\documentclass[prx,aps,twocolumn,showpacs,eqsecnum,superscriptaddress]{revtex4-1}

\usepackage{amssymb}
\usepackage{amsmath}
\usepackage{graphicx}
\usepackage{wasysym}
\usepackage{mathtools}

\newcommand{\be}{\begin{eqnarray}}
\newcommand{\ee}{\end{eqnarray}}
\newcommand{\bbm}{\begin{bmatrix}}
\newcommand{\ebm}{\end{bmatrix}}
\newcommand{\bpm}{\begin{pmatrix}}
\newcommand{\epm}{\end{pmatrix}}
\renewcommand{\v}[1]{{\bf #1}}

\newcommand{\parallelsum}{\mathbin{\!/\mkern-5mu/\!}}

\begin{document}
\title[]{Classification of flat bands according to the band-crossing singularity of Bloch wave functions}

\author{Jun-Won \surname{Rhim}}
\affiliation{Department of Physics and Astronomy, Seoul National University, Seoul 08826, Korea}
\affiliation{Center for Correlated Electron Systems, Institute for Basic Science (IBS), Seoul 08826, Korea}

\author{Bohm-Jung \surname{Yang}}
\email{bjyang@snu.ac.kr} 
\affiliation{Department of Physics and Astronomy, Seoul National University, Seoul 08826, Korea}
\affiliation{Center for Correlated Electron Systems, Institute for Basic Science (IBS), Seoul 08826, Korea}
\affiliation{Center for Theoretical Physics (CTP), Seoul National University, Seoul 08826, Korea}

\begin{abstract}
We show that flat bands can be categorized into two distinct classes, that is, singular and nonsingular flat bands, by exploiting the singular behavior of their Bloch wave functions in momentum space. In the case of a singular flat band, its Bloch wave function possesses immovable discontinuities generated by the band-crossing with other bands, and thus the vector bundle associated with the flat band cannot be defined. This singularity precludes the compact localized states from forming a complete set spanning the flat band. Once the degeneracy at the band crossing point is lifted, the singular flat band becomes dispersive and can acquire a finite Chern number in general, suggesting a new route for obtaining a nearly flat Chern band. On the other hand, the Bloch wave function of a nonsingular flat band has no singularity, and thus forms a vector bundle. A nonsingular flat band can be completely isolated from other bands while preserving the perfect flatness. All one-dimensional flat bands belong to the nonsingular class. We show that a singular flat band displays a novel bulk-boundary correspondence such that the presence of the robust boundary mode is guaranteed by the singularity of the Bloch wave function. Moreover, we develop a general scheme to construct a flat band model Hamiltonian in which one can freely design its singular or nonsingular nature. Finally, we propose a general formula for the compact localized state spanning the flat band, which can be easily implemented in numerics and offer a basis set useful in analyzing correlation effects in flat bands.
\end{abstract}



\maketitle


\section{Introduction}
A flat band, strictly dispersionless in the whole Brillouin zone~\cite{Sutherland1986,Aoki1993,Matsumura1996,Tasaki1998,Yoshino2004,Balents2008,Vanmaekelbergh2014,Desyatnikov2014,Schmidt2015,Lai2015,Flach2016,Vicencio2016,Flach2017o,Flach2017c,Aoki2017,Flach2018r,Schmelcher2018}, has been considered as an ideal playground to explore strong correlation physics due to the complete quenching of the kinetic energy~\cite{Lieb1989,Sarma2007,Cooper2012,Zhang2013,Hausler2015,Derzhko2015,Volovik2016,Torma2016,Nguyen2018,Li2018,Wen2011,Sarma2011,Mudry2011,Sheng2011,Bernevig2011,Ran2011,Franz2012,Bergholtz2012,Sarma2012,Lauchli2012,Liu2013a,Liu2013b,Udagawa2017,TaoLi2018,Tovmasyan2016}.
For example, a number of intriguing theoretical predictions are proposed in flat band systems such as the Wigner crystallization in the honeycomb lattice~\cite{Sarma2007}, the nontrivial conductivity behavior in the presence of long-range Coulomb interactions~\cite{Hausler2015}, and the huge critical temperature for the superconductivity~\cite{Volovik2016}.
Also, a nearly flat band with a finite Chern number  was recently proposed as a promising platform to realize fractional Chern insulators, analogous to the case of the flat Landau level~\cite{Wen2011,Sarma2011,Mudry2011,Sheng2011,Bernevig2011,Ran2011,Franz2012,Bergholtz2012,Sarma2012,Lauchli2012,Liu2013a,Liu2013b}.

Up to now, several flat band models have been experimentally realized in the photonic crystals~\cite{Thomson2015,Amo2018, Chen2016a, Chen2016b}, optical lattices~\cite{Manninen2013,Takahashi2015,Bloch2016,Takahashi2017}, manipulated atomic lattices~\cite{Liljeroth2017,Swart2017}, and various metamaterials~\cite{Kitano2012,Kitano2016}.
For instance, in photonic systems, a flat band has been considered as a promising route to realize slowly propagating light~\cite{Baba2008}.  
Interestingly, the experimental observation of nearly flat bands was also reported even in conventional solid state systems recently. 
For example, in the twisted bilayer graphene at magic angle, it is proposed that the presence of almost flat bands is the fundamental origin of the Mott insulating phases and the associated superconductivity~\cite{Herrero2018a,Herrero2018b}.
Also, in the layered Fe$_3$Sn$_2$, a nearly dispersionless band is detected by ARPES measurements~\cite{Zhang2018}.

The localized nature of the flat band is usually captured by strictly localized eigenfunctions in real space, so-called the compact localized state (CLS)~\cite{Sutherland1986,Matsumura1996,Flach2017o,Dubail2015,Read2017}.
The CLS can be considered as an extreme limit of the Wannier function whose amplitude is finite only in a bounded region in real space, and completely vanishes outside of it.
Such a compact localization is possible because of the destructive interferences between the wave function amplitudes after the hopping processes of the Hamiltonian, and in many cases this phenomena originate from the specific lattice structures supporting geometric frustration.
Because of this, most of the previous studies on flat bands have paid attention to particular lattice structures, and the understanding of the universal properties of flat bands, which are independent of the detailed lattice structure or spatial dimensionality, is quite limited. 
In particular, considering that a perfectly flat band isolated from other bands has a zero Chern number, it is generally believed that the band topology of flat bands is trivial in momentum space~\cite{Tang2014}.
While a completely flat Landau level obtained from a usual continuum model has a nonzero Chern number, we focus on the lattice models with finite hopping range.

Interestingly, however, a recent theoretical study of itinerant electron models in frustrated lattices has reported intriguing momentum space structures of flat band systems.
For instance, it is found that a flat band in the kagome lattice exhibits a band crossing with another dispersive band at a particular momentum.
Bergman \textit{et al} have pointed out that such a band degeneracy is related with the incompleteness of the CLSs in this geometrically frustrated system~\cite{Balents2008}.
Namely, the full set of CLSs including all CLSs connected by lattice translation vectors are found to be linearly dependent to each other, and the missing basis eigenstates should be complemented by the so-called non-contractible loop states (NLSs) which are compact-localized in one direction but extended in the other direction.
This suggests that the flat band possessing a band crossing might be distinguished from other types of flat bands topologically, because NLSs cannot be smoothly deformed to CLSs in real space on the torus geometry respecting the periodic boundary condition~\cite{note1}.

Here we show that the universal properties of flat band systems can be described in a unified way by investigating a certain singular property of the Bloch wave functions in momentum space.
This is quite an unexpected outcome considering that only local symmetries of a given lattice model conventionally have been considered to study and generate flat bands~\cite{Matsumura1996,Desyatnikov2014,Flach2017o,Vicencio2016,Aoki2017}.
We show that the absence of the complete set of the CLSs for a flat band is related to the existence of the \textit{immovable discontinuity} of the Bloch wave function in the Brillouin zone.
The term \textit{immovable discontinuity} stands for the nonexistence of the local gauge choice that makes the Bloch wave function continuous around a certain momentum by shifting the position of the singular point~\cite{Strinati1977}.
The presence of the immovable discontinuity in the flat band, which is the defining property of a singular flat band, implies that it is touching with another dispersive band at the singular point, which we call \textit{a singular band touching}.
This kind of the singularity is distinct from that of a Chern band.
In the Chern band case, one can always make a local gauge choice shifting the location of the singularity to another momentum~\cite{Dubail2015,Kohmoto}.
On the other hand, even if a flat band is touching with another band, in some cases, one can choose a gauge in which the wave function is continuous. 
In this case, the band touching point is called \textit{a non-singular touching}.
The flat band with a non-singular band touching can be spanned by a complete set of CLSs as in the case when the flat band is fully separated from other bands.
%

%
Let us note that Dubail and Read also studied the CLS from the perspective of the Bloch wave functions\cite{Dubail2015}, and N. Read classified the non-singular flat bands by applying the algebraic K-theory\cite{Read2017}. While these two works considered the cases where the vector bundle is well-defined due to the energy gap between the filled and unfilled bands, we have focused on the opposite cases where the vector bundle associated with the flat band cannot be defined due to the singular band touching with other dispersive bands.


These two types of band touching in flat band systems display completely different features when the degeneracy at the crossing point is lifted. 
In the case of a non-singular band touching, one can always open the gap while preserving the band flatness, and the resulting isolated flat band is topologically trivial. 
On the other hand, a singular flat band always becomes dispersive after gap-opening, which can lead to a nearly flat band with a finite Chern number. 
This process provides a new scheme to obtain a nearly flat Chern band starting from a singular flat band.
This property clearly demonstrates that the nature of a band crossing in flat band systems is strongly constrained by the discontinuity of the Bloch wave function, which, in turn, critically affects the band flatness and its topological nature after degeneracy lifting.
Although there is no definite local topological invariant, such as a winding number, characterizing the band crossing point in flat band systems, the singularity of the Bloch states manifests non-trivially combined with the band flatness condition.

Furthermore, we show in general that this singularity is manifested in real space as localized boundary modes of an open geometry whose penetration depth is smaller than the size of the CLSs.
These boundary states are actually precursors of the NLSs in 2D and the non-contractible planar states (NPSs) in 3D systems with the periodic boundary condition.
We also discuss how to probe this boundary mode experimentally.

Finally, we propose several general and practical schemes for tailoring CLSs and flat band tight binding models. 
Up to now, CLSs have been constructed based on some intuition, which works only for limited simple models. 
The scheme we developed, however, is so general that CLSs can be easily constructed even for complex systems and one can even freely determine the singular or non-singular nature of the flat band in a controlled manner.

\section{Discontinuities of the Bloch wave function and incompleteness of the compact localized states}\label{sec:cls}

We study the properties of the CLSㄴ from the perspective of the Bloch wave function.
First, we show rigorously that if there exists a flat band, one can always find a set of CLSs as degenerate eigenstates whose energy is the same as that of the flat band. 
This holds regardless of the dimensionality, the lattice structure, and the presence or absence of the band touching between the flat band and other bands.
When the system is composed of \textit{N} unit cells, \textit{N} independent CLSs are necessary to span a flat band completely. 
Below we show that such a complete set of CLSs does not exists if the Bloch wave function associated with the flat band possesses an immovable discontinuity in momentum space due to the band touching.

\subsection{The existence of the compact localized state}\label{sec:cls_existence}

The eigenfunction of a Bloch Hamiltonian $\mathcal{H}_\mathbf{k}$ can generally be written as

\begin{align}
|\psi_{n,\mathbf{k}}\rangle = \frac{1}{\sqrt{N}}\sum_{\mathbf{R}}\sum_{q = 1}^Q e^{i\mathbf{k}\cdot\mathbf{R}} v_{n,\mathbf{k},q} |\mathbf{R},q\rangle,
\end{align}
where $n$ is the band index and $\mathbf{R} = \sum_{l=1}^d m_l \mathbf{a}_l$ is the lattice vector for the $d$ dimensional system consisting of $N$ unit cells.
$m_l$ is an integer, and $\mathbf{a}_l$ is the primitive vector.
$v_{n,\mathbf{k},q}$ is the $q$-th component of the column vector $\v v_{n,k}$ which is the eigenvector of $\mathcal{H}_{\v k}$ with energy $\epsilon_{n,\mathbf{k}}$.
The number of components of $\v v_{n,k}$ is identical to the number of sites and orbitals in the unit cell. 
$|\mathbf{R},q\rangle = c^\dag_{\mathbf{R},q}|0\rangle$ where $c^\dag_{\mathbf{R},q}$ is an operator creating an electron in the $q$-th orbital in the unit cell at $\v R$ and $|0\rangle$ indicates the vacuum state.
We assume $\mathcal{H}_\mathbf{k}$ is a $Q\times Q$ matrix.
Here, we assign the same Bloch phase $e^{i\mathbf{k}\cdot\mathbf{R}}$ to all the orbitals in the same unit cell so that $\mathcal{H}_\mathbf{k} = \mathcal{H}_{\mathbf{k}+\mathbf{G}}$ where $\mathbf{G} = \sum_{l=1}^d m_l \mathbf{b}_l$ is the reciprocal lattice vector with $\mathbf{b}_l$ the primitive reciprocal vector satisfying $\mathbf{a}_{l}\cdot\mathbf{b}_{l^\prime} = 2\pi\delta_{l,l^\prime}$.

We consider the case where there is at least one flat band, and focus on one flat band while omitting its band index for simplicity from now on.
Since all the Bloch eigenfunctions in the same flat band are degenerate, one can freely mix them to obtain a new eigenfunction as follows.
\begin{align}
|\chi_\mathbf{R}\rangle &= c_\chi \sum_{\mathbf{k}\in \mathrm{BZ}} \alpha_\mathbf{k} e^{-i\mathbf{k}\cdot\mathbf{R}} |\psi_{\mathbf{k}}\rangle = \sum_{\mathbf{R}^\prime}\sum_{p=1}^Q \mathbf{A}_{\mathbf{R},\mathbf{R}^\prime,q} \cdot |\mathbf{R}^\prime,q\rangle, \label{eq:cls_0}
\end{align}
where $\mathbf{A}_{\mathbf{R},\mathbf{R}^\prime,q}$, the $q$-th component of the column vector
\begin{align}
\mathbf{A}_{\mathbf{R},\mathbf{R}^\prime} = \frac{c_\chi}{\sqrt{N}}\sum_{\mathbf{k}\in \mathrm{BZ}}\alpha_\mathbf{k} \exp \left[ i\mathbf{k}\cdot(\mathbf{R}^\prime-\mathbf{R})\right] \mathbf{v}_{\mathbf{k}}, \label{eq:A}
\end{align}
estimates the amplitudes of the wave function in the unit cell at $\mathbf{R}^\prime$. 
$c_\chi$ is the normalization constant.
If there exists a scalar function $\alpha_\mathbf{k}$ that makes $\mathbf{A}_{\mathbf{R},\mathbf{R}^\prime}$ nonzero only in a certain finite region, we call $|\chi_\mathbf{R}\rangle$ a CLS.
Once we obtain a CLS around $\mathbf{R}$, any translated copies of it $|\chi_{\mathbf{R}-\mathbf{R}_0}\rangle$ are also eigenstates.
In this way, one can find a set of $N$ different CLSs.

For compact localization, each component of $\alpha_\mathbf{k} \mathbf{v}_{\mathbf{k}}$ should be a finite sum of the Bloch phases (FSBP) since $\mathbf{A}_{\mathbf{R},\mathbf{R}^\prime}$ is just an inverse Fourier transformation of $\alpha_\mathbf{k} \mathbf{v}_{\mathbf{k}}$.
That is,
\begin{align}
\alpha_\mathbf{k}v_{\mathbf{k},q} = \sum_{m_1,\cdots,m_d} f_{m_1,\cdots,m_d}^{(q)} \exp\left(i\sum_{l=1}^d m_l\mathbf{k}_l\cdot\mathbf{a}_l\right), \label{eq:fsbp}
\end{align}
where $v_{\v k,q}$ is the $q$-th component of the column matrix $\v v_{\v k}$, $m_l$ runs from $m_l^{(\mathrm{lo})}$ to $m_l^{(\mathrm{up})}$, and $f_{m_1,\cdots,m_d}^{(q)}$ is a complex number.
Due to the upper and lower limits of $m_l$, the CLS's coefficient $\mathbf{A}_{\mathbf{R},\mathbf{R}^\prime}$ vanishes if one of $m_i$'s in $\mathbf{R}^\prime - \mathbf{R} = \sum_i m_i\mathbf{a}_i$ is out of this range.
In other words, $\alpha_\mathbf{k}v_{\mathbf{k},q}$ is a finite polynomial of $X_l = \exp(i\mathbf{k}_l\cdot\mathbf{a}_l)$.

\begin{figure}
	\begin{center}
		\includegraphics[width=1\columnwidth]{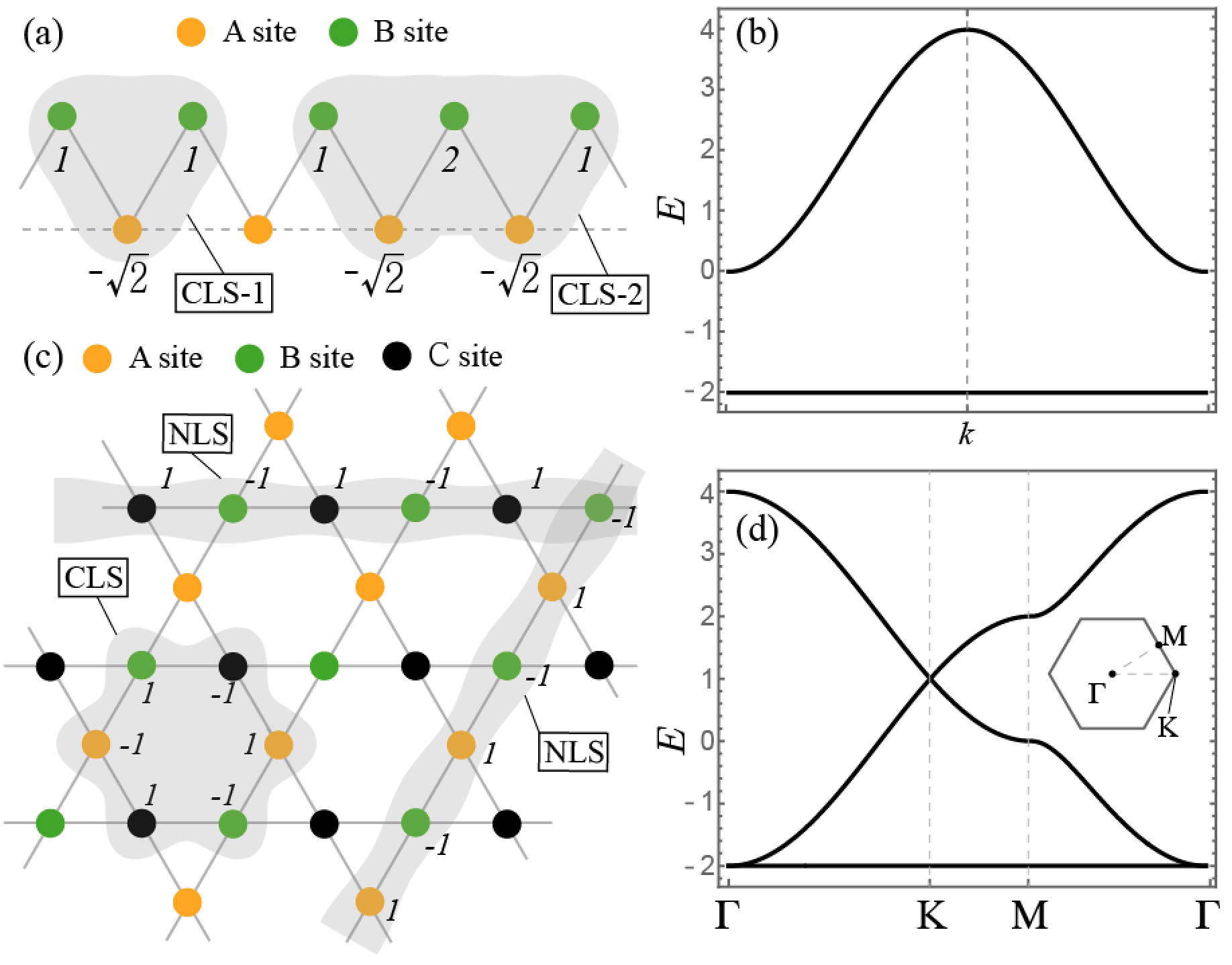}
	\end{center}
	\caption{ (a) 1D flat band model on a zigzag chain. There are two sites, A and B, per unit cell, and each site has one orbital. The solid and dashed lines represent different hopping parameters. In the zigzag chain, the hopping parameter is $V_1$ along the solid line and $V_2$ along the dashed one. There exists a flat band at $E=-2$ when $V_1=\sqrt{2}$ and $V_2=1$. Two examples of the CLS for this case are shown by the gray compact regions. The numbers near the sites in this region represent the amplitudes of the CLS. (b) The band structure for $V_1=\sqrt{2}$ and $V_2=1$. (c) The kagome lattice with the nearest neighboring hopping processes. Three colors of the sites stand for the three different basis sites. The simplest CLS and two independent NLSs are exhibited by gray regions. (d) The band structure for the kagome lattice is drawn between high symmetry points.
	}
	\label{fig:zigzag}
\end{figure}

For a flat band, such $\alpha_{\v k}$ that makes the vector $\alpha_\mathbf{k} \mathbf{v}_{\mathbf{k}}$ in the form of the FSBP always exists if a given Hamiltonian contains only finite-range hopping processes.
Since $\mathbf{v}_{\mathbf{k}}$ is an eigenvector of the Hamiltonian, $\alpha_\mathbf{k} \mathbf{v}_{\mathbf{k}}$ is just a unnormalized eigenvector which is a solution of
\begin{align}
\bar{\mathcal{H}}_\mathbf{k} \mathbf{x}_{\mathbf{k}} = \left( \mathcal{H}_\mathbf{k}-\epsilon_0\mathcal{I} \right) \mathbf{x}_{\mathbf{k}} = 0,
\end{align}
where $\epsilon_0$ is the flat band's energy, and $\v x_{\v k} \propto \alpha_\mathbf{k} \mathbf{v}_{\mathbf{k}}$.
Since $\epsilon_0$ is a constant and all the elements of $\mathcal{H}_\mathbf{k}$ are in the form of the FSBP. 
This system of homogeneous equations can also have a solution in the form of the FSBP which leads to the conclusion that such $\alpha_{\v k}$ is guaranteed.
We note that the same conclusion was derived by N. Read by applying the algebraic K-theory\cite{Read2017}.
A more rigorous proof for this is given in App.~\ref{app:cls_existence}.

\subsection{The completeness condition for the CLSs}

Once a CLS of a given flat band is found, its lattice translations give $(N-1)$ different copies of CLSs, and these $N$ CLSs are expected to span the flat band completely.
This may explain why the electrons in the flat band are immobile even though there are hopping processes.
However, the linear independence of those $N$ translated copies of CLSs is not guaranteed in general.
In this section, we derive the exact condition for the completeness of the $N$ translated copies of CLS from the perspective of the Bloch wave function by analyzing its discontinuities in momentum space.

The completeness of the $N$ translated copies of the CLS can be examined by using its expression in (\ref{eq:cls_0}) as follows:
\begin{align}
D =  \begin{vmatrix}
 \alpha_{\mathbf{k}_1}e^{-i\mathbf{k}_1\cdot\mathbf{R}_1}  & \cdots & \alpha_{\mathbf{k}_1}e^{-i\mathbf{k}_1\cdot\mathbf{R}_N} \\
 \alpha_{\mathbf{k}_2}e^{-i\mathbf{k}_2\cdot\mathbf{R}_1}  & \cdots & \alpha_{\mathbf{k}_1}e^{-i\mathbf{k}_2\cdot\mathbf{R}_N} \\
 \vdots  & \ddots & \vdots \\
 \alpha_{\mathbf{k}_N}e^{-i\mathbf{k}_N\cdot\mathbf{R}_1}  & \cdots & \alpha_{\mathbf{k}_N}e^{-i\mathbf{k}_N\cdot\mathbf{R}_N}
 \end{vmatrix} \propto \prod_{l=1}^N\alpha_{\mathbf{k}_l} .\label{eq:det_1}
\end{align}
Here, we use the fact that the set of Bloch wave functions 
$\{ \psi_{\mathbf{k}_l} \}$ is a complete basis for the flat band.
Each column in the determinant of (\ref{eq:det_1}) represents the vector corresponding to $|\chi_{\mathbf{R}_l}\rangle$ in this basis.
From (\ref{eq:det_1}), we obtain the most basic conclusion that \textit{if $\alpha_\mathbf{k}$ is nonzero at every $\v k$, the $N$ translated copies of the CLS obtained from $\alpha_\mathbf{k}$ form a complete set.}
We call this kind of flat band a \textit{non-singular} flat band.
On the other hand, if any possible $\alpha_\mathbf{k}$ vanishes at a momentum, the corresponding flat band is called a \textit{singular} flat band.

Let us consider a 1D zigzag lattice, illustrated in Fig.~\ref{fig:zigzag}(a), as an example.
It has two sites in the unit cell, and there is one orbital per site~\cite{Zhang2015}.
Considering the hopping amplitudes between nearest-neighbors $V_1$ and second nearest-neighbors $V_2=V_1/\sqrt{2}$, the Hamiltonian is given by
\begin{align}
\mathcal{H}_{k} = \bpm \sqrt{2}V_1\cos k & V_1+V_1e^{-ik} \\ V_1+V_1e^{ik} & 0\epm,
\end{align}
where $\mathcal{H}_{\v k}|_{11}$ and $\mathcal{H}_{\v k}|_{22}$ correspond to the on-site potentials on A and B sites, respectively.
For the flat band at $E=-\sqrt{2}V_1$, the eigenvector is given by
\begin{align}
\mathbf{v}_{k}^\mathrm{flat} = \frac{1}{\sqrt{4+2\cos k}}\bpm  -\sqrt{2} \\ 1+e^{ik} \epm.
\end{align}
One can easily find that $\alpha_\mathbf{k} = \sqrt{4+2\cos k}$ makes $\alpha_\mathbf{k}\mathbf{v}_{\mathrm{zigzag},k}^\mathrm{flat}$ in the form of the FSBP.
Then, from (\ref{eq:A}), the relevant CLS is given by
\begin{align}
\mathbf{A}_{0,R} =& \frac{1}{2}\bpm -\sqrt{2}\delta_{R,0} \\ \delta_{R,0} + \delta_{R,-1} \epm
\end{align}
around the unit cell at $R=0$.
The amplitude at the A site of the unit cell at $R=0$ is $-1/\sqrt{2}$, and those at two nearest neighboring B sites are $1/2$. 
This CLS is illustrated in Fig.~\ref{fig:zigzag}(a).
Due to translational invariance, one can find $N$ CLSs centered at different unit cells.
Since $\alpha_k =\sqrt{4+2\cos k}$ is always nonzero in the Brillouin zone, those CLSs form a complete set, and the flat band is nonsingular completely described by the CLSs.
So, we call the flat band of the 1D zigzag lattice as a non-singular flat band.

On the other hand, with another choice $\tilde{\alpha}_k = \sqrt{4+2\cos k}(1+e^{-ik})$, the relevant CLS is of the form
\begin{align}
\tilde{\mathbf{A}}_{0,R} = \bpm -\sqrt{2}\left(\delta_{R,0} + \delta_{R,1}\right)\\ 2\delta_{R,0} + \delta_{R,1} + \delta_{R,-1} \epm,
\end{align}
which is shown in Fig.~\ref{fig:zigzag}(a).
Since $\tilde{\alpha}_k$ vanishes at $k=\pi$, the $N$ copies of CLSs obtained after translating $\tilde{\mathbf{A}}_{0,R}$ do not form a complete set spanning the 1D chain with a ring geometry.
Let us note that, in this case, the completeness of the CLSs actually depends on whether the system size $N$ is even or odd because the value of $(2\pi/N)m$, which is the momentum under the periodic boundary condition, can be strictly $\pi$ only when $N$ is even.
This is explicitly shown by the fact that $\sum_{s=0}^{N-1} (-1)^s\tilde{\mathbf{A}}_{s,R} = 0$ for even $N$ case whereas one cannot find such a constraint for odd $N$ case.
This means that even in the non-singular flat band, we can find a choice of $\alpha_{\v k}$ that makes the resulting CLSs linearly dependent.
However, a given flat band is non-singular if there exist at least one choice of $\alpha_{\v k}$ which is nonzero for all $\v k$.

\subsection{Immovable discontinuity and the incompleteness of the CLSs}\label{sec:irremovable_discon}

As an opposite situation to the previous subsection, if every choice of $\alpha_\mathbf{k}$ vanishes at a momentum $\v k$, one cannot span the flat band only with the CLSs because the determinant (\ref{eq:det_1}) vanishes.
However, it is impossible to find all the possible forms of $\alpha_\mathbf{k}$.
To resolve this problem, we show that the zeros of $\alpha_\mathbf{k}$ is closely related to the immovable discontinuities of the Bloch wave function $\v v_{\v k}$ of the flat band as follows.
%
%
\textit{If there exists a nonzero function $\alpha_\mathbf{k}$ which makes $\alpha_\mathbf{k}\mathbf{v}_\mathbf{k}$ a FSBP, the eigenvector $\mathbf{v}_\mathbf{k}$ can be made continuous (non-singular) by the proper gauge choice}.
Or, equivalently, \textit{if $\mathbf{v}_\mathbf{k}$ is always discontinuous at some $\mathbf{k}_0$ for any local gauge choice around it, any $\alpha_\mathbf{k}$ should be vanishing at $\mathbf{k}_0$, that is, any kind of the $N$ translated copies of the CLSs cannot span the flat band.}
Below, we justify this statement.

According to the previous section, the eigenvector of a flat band can be chosen to be proportional to $\bpm w_{\mathbf{k},1} & \cdots & w_{\mathbf{k},Q} \epm^\mathrm{T}$ where $w_{\mathbf{k},q}$ is the complex function in the form of the FSBP like $\alpha_{\v k}\v v_{\v k}$ in (\ref{eq:fsbp}), and $Q$ is the size of the Hamiltonian matrix.
Without loss of generality, one can assume that $w_{\mathbf{k},q}$'s have no momentum-dependent common factor.
Then, the normalized eigenvector is of the form
\begin{align}
\mathbf{v}_\mathbf{k} = \frac{1}{\sqrt{\sum_{q=1}^Q |w_{\mathbf{k},q}|^2}}\bpm w_{\mathbf{k},1} \\ \vdots \\  w_{\mathbf{k},Q} \epm. \label{eq:eigenvector}
\end{align}
If the eigenvector is discontinuous at $\mathbf{k}_0$, every $w_{\mathbf{k},q}$ has to be vanishing at that momentum.
Otherwise, the denominator $(\sum_{q=1}^Q |w_{\mathbf{k},q}|^2)^{1/2}$ cannot be zero and all the components of the eigenvector are continuous since any function composed of the FSBP is continuous.
Therefore $\alpha_\mathbf{k}$, which must be proportional to $(\sum_{q=1}^Q |w_{\mathbf{k},q}|^2)^{1/2}$, is zero at the momentum $\v k = \v k_0$.
This proves the general statement in the previous paragraph.

One can understand the nature of the immovable discontinuity in the above by comparing it with the singularity of the conventional Chern band as follows.
%
%
The discontinuity of $\v v_{\v k}$ at $k_0$ appears since the value of $\v v_{\v k}$ varies depending on the path along which $\v k$ approaches $\v k_0$.
That is, $\v v_{\v k}$ is a discontinuous function of $\v k$ for which partial derivatives exist.
The necessary condition for this discontinuity is the band touching or degeneracy at $\v k_0$ which cannot be gauged away because the band touching itself is gauge independent.
This is why we call such a discontinuity immovable.
Due to the discontinuity of the Bloch wave function, the flat band does not form a vector bundle, and thus the Chern number cannot be defined\cite{Dubail2015}.
The Chern band also has the singularity in its Bloch wave function.
However, in this case, the Bloch wave function forms a vector bundle because one can shift the singularity to another $\v k$ point.
Using this property, one can prepare a number of patches consisting of analytic vector bundles to cover the whole momentum space\cite{Kohmoto}.
In this sense, the singularity of the Chern band is movable.

The relation between the incompleteness of $N$ translated CLSs and the zeros of $\alpha_{\v k}$ can be understood more easily from the very first expression of the CLS (\ref{eq:cls_0}).
Since $\alpha_\mathbf{k}$ is the coefficient of the Bloch wave function $|\psi_{\v k}\rangle$, the Bloch wave function at $\v k_0$ does not participate in the construction of the CLS.
This implies that the number of linearly independent eigenvectors among $N$ translated CLSs is less than $N$ and we should add some \textit{non-compact} or extended states to span the flat band completely.
These non-compact states will be further discussed in detail in Sec.~\ref{sec:bbc}.
Note that our conclusion holds not only for the $N$ translated copies of CLSs but also for the general set of $N$ CLSs with different shapes because any form of $\alpha_{\v k}$ vanishes at the singular momentum $\v k_0$.

As an example, let us consider the following Hamiltonian describing nearest neighbor hopping on the kagome lattice,
\begin{align}
\mathcal{H}_\mathbf{k} = -t\bpm 0 & e^{-i\mathbf{a}_3\cdot\mathbf{k}}+1 & e^{i\mathbf{a}_2\cdot\mathbf{k}}+1 \\ e^{i\mathbf{a}_3\cdot\mathbf{k}}+1 & 0 & e^{-i\mathbf{a}_1\cdot\mathbf{k}}+1 \\ e^{-i\mathbf{a}_2\cdot\mathbf{k}}+1 & e^{i\mathbf{a}_1\cdot\mathbf{k}}+1 & 0\epm,\label{eq:kagome_ham}
\end{align}
where $t$ is the nearest neighbor hopping parameter, and $\mathbf{a}_1 = (1,0)$, $\mathbf{a}_2 = (-1/2,\sqrt{3}/2)$, and $\mathbf{a}_3 = -\mathbf{a}_1 - \mathbf{a}_2$.
There is a flat band at $E=2t$ with the eigenvector 
\begin{align}
\mathbf{v}_\mathbf{k} = c_\mathbf{k}\bpm e^{i\v a_1\cdot\v k}-1 \\ 1 - e^{-i\v a_2\cdot\v k}  \\ e^{-i\v a_2\cdot\v k} - e^{i\v a_1\cdot\v k} \epm,
\end{align}
where $c_\mathbf{k} = \{2 ( 3-\cos k_x -2\cos k_x/2\cos \sqrt{3}k_y/2 )\}^{-1/2}$.
To make CLSs, we choose $\alpha_\mathbf{k}=c_\mathbf{k}^{-1}$.
This leads to
\begin{align}
\v A_{0,R} = \frac{1}{\sqrt{6}}\bpm \delta_{\v R,-\v a_1} - \delta_{\v R,0} \\ \delta_{\v R,0} - \delta_{\v R,\v a_2} \\ \delta_{\v R,\v a_2} - \delta_{\v R,-\v a_1} \epm, \label{eq:cls_kagome}
\end{align}
which is illustrated in Fig.~\ref{fig:zigzag}(c).
Let us note that the $N$ translated copies of $\v A_{0,R}$ are not linearly independent of each other due to the discontinuity of $\mathbf{v}_\mathbf{k}$ at $\v k=0$, that is, the value of $\lim_{\mathbf{k}\rightarrow 0}\mathbf{v}_\mathbf{k}$ depends on how we approach the $\Gamma$ point.
For example, $\lim_{k_x\rightarrow 0}\mathbf{v}_{(k_x,0)} \neq \lim_{k_y\rightarrow 0}\mathbf{v}_{(0,k_y)}$.
This is reflected in the fact that the $\alpha_\mathbf{k}=1/c_{\v k}$ vanishes at $\mathbf{k}=0$.
Actually, the sum of the $N$ copies of CLSs vanishes under the periodic boundary condition in which the system has a torus geometry.~\cite{Balents2008}.
Since any other choice of $\alpha_\mathbf{k}$ should be proportional to $c_\mathbf{k}^{-1}$, any possible form of the CLS cannot span a complete set.
In the kagome lattice case, the incompleteness does not depend on the system size unlike the 1D zigzag lattice case discussed in the previous subsection because the momentum $\mathbf{k}=0$ is always allowed on the torus geometry of the system.
As a result, the flat band cannot be completely described by the CLSs, and some extended states, called the NLSs~\cite{Balents2008} as illustrated in Fig.~\ref{fig:zigzag}(c), must be complemented in addition to the CLSs.
As noted from this example, the equivalence between the wave function's discontinuity and the absence of nonzero $\alpha_\mathbf{k}$ offers an extremely convenient way of determining the completeness of the $N$ translated copies of any possible CLSs.

An interesting conclusion from the general statement is that the complete set of CLSs can always be found in 1D.
That is, any flat band in 1D system is non-singular (trivial).
In 1D, the Bloch phase can be represented as $e^{ink} = z^n$ where $z=e^{ik}$, and each component $w_{k,q}$ in (\ref{eq:eigenvector}) is just a Laurent series of $z$ around $z=0$ with upper and lower limit in the power of $z$ because $w_{k,q}$  is in the form of the FSBP.
One can freely transform $w_{k,q}$ of the given eigenvector into a form of a Taylor series by multiplying the inverse of the Bloch phase with the lowest negative power of $z$.
Then, the resulting $\tilde{w}_{k,q}$ is just a finite polynomial.
In this 1D case, it is impossible for all $\tilde{w}_{k,q}$$'s$ to vanish simultaneously, for example, at $z_0$ because it means all those components should be proportional to $z-z_0$.
This implies that those components have a common factor which contradicts the original assumption that $\tilde{w}_{k,q}$$'s$ have no common factor.
So, we can always obtain a non-singular $\alpha_k$ in 1D by finding $\tilde{w}_{k,q}$$'s$ without the common factor, and the $N$ translated copies of a CLS span the flat band completely.

\begin{figure}
	\begin{center}
		\includegraphics[width=1\columnwidth]{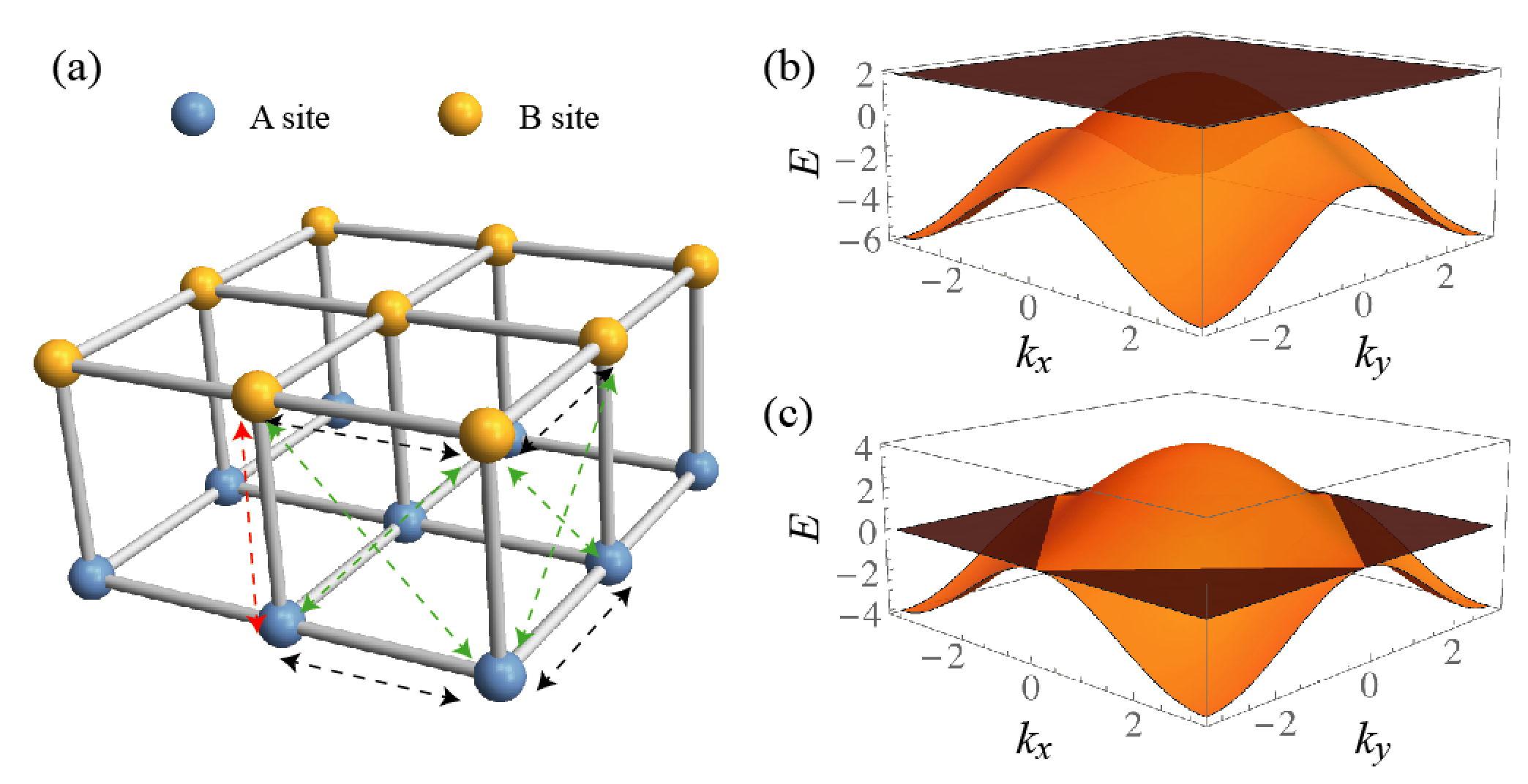}
	\end{center}
	\caption{ (a) The bilayer square lattice with the nearest and next nearest neighbor hopping processes represented by dashed lines. Hopping parameters for black and green dashed lines are $1/2$ and those for red ones are $\beta$. (b) and (c) are the band structures for $\beta=-2$ and $\beta=0$.
	}
	\label{fig:nonsingular}
\end{figure}

While the discontinuity of the eigenvector comes from the band touching, not all the band touchings are singular.
Namely, even though a flat band touches other bands, it can be spanned by a set of CLSs completely if the eigenvectors of the flat band do not have any immovable singularity.
A band touching can be identified to be \textit{singular} or \textit{non-singular}, depending on  the presence or absence of the discontinuity of the corresponding eigenfunctions.
An example of a non-singular band touching appears in the bilayer square lattice model illustrated in Fig. \ref{fig:nonsingular}(a).
The corresponding Hamiltonian is given by
\begin{align}
\mathcal{H} = \bpm \cos k_x + \cos k_y & \cos k_x + \cos k_y-2 \\ \cos k_x + \cos k_y-2 & \cos k_x + \cos k_y \epm,
\end{align}
which has two eigenenergies $E_1(\mathbf{k}) = 2$ and $E_2(\mathbf{k}) = -2 + 2\cos k_x + 2\cos k_y$ as shown Fig. \ref{fig:nonsingular}(b).
Although these two bands touch each other at $k_x = k_y = 0$ quadratically, the eigenvector of the flat band, $\mathbf{v}_\mathbf{k} = (1/\sqrt{2})\bpm 1 & 1\epm^\mathrm{T}$, is non-singular.
As a result, the relevant CLSs can span the flat band completely without resorting to extended states such as NLSs.

\section{Singular quadratic band touching preserved by band flatness}\label{sec:singular_band_touching}

In this section, mainly focusing on the flat band with a quadratic band touching, we discuss how to distinguish the non-singular and singular band touchings from the modulation of the band structure when the band degeneracy at the crossing point is lifted.
We show that, in the case of a non-singular band touching, the degeneracy can be lifted while maintaining the flatness of the flat band. 
On the other hand, in the case of a singular band touching, the degeneracy lifting generally accompanies the warping of the flat band, and the resulting warped band can have a finite Chern number.

\begin{figure}
	\begin{center}
		\includegraphics[width=0.8\columnwidth]{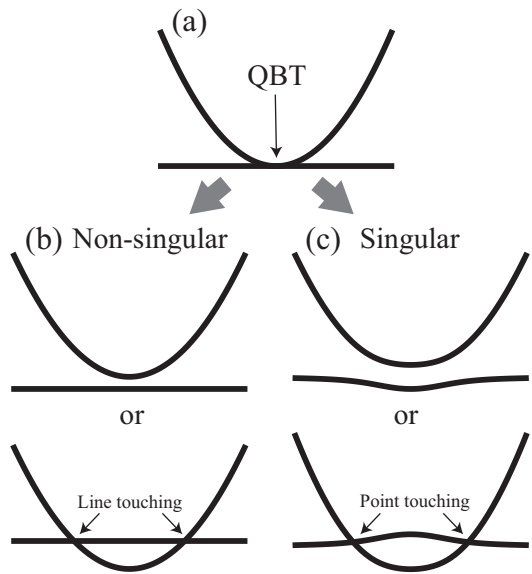}
	\end{center}
	\caption{ If the quadratic band touching in (a) is non-singular, a shifting of the flat band, upward or downward, is possible while preserving the band flatness as shown in (b). On the other hand, if it is singular, the flat band generically becomes dispersive after the band shifting process as shown in (c).
	}
	\label{fig:gap_opening}
\end{figure}

\subsection{Non-singular band touching}

Let us consider a
 unitary operator $\mathcal{U}_\mathbf{k}$ diagonalizing the $Q\times Q$ tight binding flat band Hamiltonian $\mathcal{H}_\mathbf{k}$ into $\mathcal{H}_\mathbf{k}^{(d)} = \mathrm{diag}(\epsilon_{1,\mathbf{k}}, \cdots ,\epsilon_{Q-1,\mathbf{k}},\epsilon_0)$.
We assume that the flat band touches with another dispersive band at $\v k = \v k_0$.
The last column of $\mathcal{U}_\mathbf{k}^\dag$ is just the eigenvector $\mathbf{v}_\mathbf{k}$ of the flat band ($\mathcal{U}_\mathbf{k}|_{Q,j}^* = \mathbf{v}_\mathbf{k}|_j$) which can be transformed to the form of the FSBP by multiplying some real function $\alpha_{\v k}$ as shown in Sec.~\ref{sec:cls}.
%
%
In the diagonalized basis, let us consider a perturbation that deforms all bands except the flat one given by $\mathcal{H}^{\prime (d)}_\mathbf{k} = \mathrm{diag}(\lambda_\mathbf{k},\cdots ,\lambda_\mathbf{k} , 0)$ where $|\lambda_\mathbf{k}| \ll 1$.
In the original basis, the perturbation becomes $\mathcal{H}^{\prime }_\mathbf{k} = \mathcal{U}_\mathbf{k}^\dag \mathcal{H}^{\prime (d)}_\mathbf{k} \mathcal{U}_\mathbf{k}  = \lambda_\mathbf{k}\mathcal{I} - \mathcal{U}_\mathbf{k}^\dag\mathrm{diag}(0,\cdots ,0 ,\lambda_\mathbf{k})) \mathcal{U}_\mathbf{k}$.
Namely, the matrix element of $\mathcal{H}^{\prime }_\mathbf{k}$ is given by
\begin{align}
\mathcal{H}^{\prime }_\mathbf{k}\big|_{ij} = \lambda_\mathbf{k}\left(\delta_{ij}- \mathcal{U}_\mathbf{k}|_{Q,i}^* \mathcal{U}_\mathbf{k}|_{Q,j} \right) = \lambda_\mathbf{k}\left(\delta_{ij}- \v v_\mathbf{k}|_{i}^* \v v_\mathbf{k}|_{j} \right).
\end{align}
This implies that $\lambda_\mathbf{k} = \lambda|\alpha_\mathbf{k}|^2$ ensures all the elements of $\mathcal{H}^{\prime}_\mathbf{k}$ are in the form of the FSBP because $\alpha_{\v k}\mathbf{v}_\mathbf{k}$ and $|\alpha_{\v k}|^2$ are all in the form of the FSBP.
Here, it is important to note that the perturbation is also in the form of the FSBP like $\mathcal{H}_\mathbf{k}$ because we only consider finite-range hopping processes.
Since $\alpha_\mathbf{k}$ can be chosen to be nonzero for all $\v k$ for the non-singular case, adding $\mathcal{H}^{\prime}_\mathbf{k}$ to $\mathcal{H}_\mathbf{k}$ removes the band touching at $\v k_0$.
Then, depending on the sign of $\lambda$, we see either gap-opening or line crossings between the flat band and the dispersive band touching with it as shown in Fig.~\ref{fig:gap_opening}(b).

For instance, one can find a perturbation that can destroy the non-singular touching in the bilayer square lattice in Sec.~\ref{sec:irremovable_discon}.
With the choice $\alpha_{\v k} =1$, we have $\mathcal{H}^{\prime}_\mathbf{k} = \lambda/2\sigma_0 + \lambda/2\sigma_x$ by noting $\mathcal{U} = 1/\sqrt{2}\sigma_0 -i1/\sqrt{2}\sigma_y$.
Then, the perturbed Hamiltonian $\mathcal{H}_{\v k} + \mathcal{H}^\prime_{\v k}$ yields eigenenergies as $E_1(\mathbf{k}) = 2$ and $E_2(\mathbf{k}) = \lambda -2 + 2\cos k_x + 2\cos k_y$.
Namely, the perturbation induces a constant shift of the dispersive band by $\lambda$, which either opens a gap ($\lambda>0$) or deforms the point touching into a line touching ($\lambda<0$) as shown in Fig.~\ref{fig:nonsingular}(c).

While the above discussion is completely general, one can understand the result more concretely by considering an effective low energy continuum model around the touching.
We deal with the quadratic band touching with the flat band which we mostly encounter with.
In 1D, the generic form of the quadratic expansion of the flat band Hamiltonian around the touching point is given by $\mathcal{H} = a_x k_x^2\sigma_x + a_y k_x^2\sigma_y + a_z k_x^2\sigma_z + a_0 k_x^2\sigma_0$ where $a_0 =\pm \sqrt{a_x^2+a_y^2+a_z^2}$, and the momentum $\v k$ is measured with respect to the touching point.
This is always non-singular which is consistent with the argument in Sec.~\ref{sec:irremovable_discon}, and one can freely shift the flat band by the perturbation $\mathcal{H}^\prime = \lambda (a_x\sigma_x + a_y\sigma_y + a_z\sigma_z +a_0\sigma_0 )$.

In 2D, as shown in detail in App.~\ref{app:low_energy}, the effective low energy Hamiltonian for the non-singular quadratic touching can always be transformed to
\begin{align}
\mathcal{H}_{\v k} = (t_1^\prime k_x^2 + t_2^\prime k_x k_y + t_3^\prime k_y^2)(\sigma_z+\sigma_0),
\end{align}
where the relevant eigenvectors, $\bpm 1 & 0 \epm$ and $\bpm 0 & 1 \epm$, are obviously non-singular at all momenta.
In this case, since there is only one species of the Pauli matrix, one can always find the perturbation of the form $\delta \sigma_z$ which lifts the double degeneracy at $\v k =0$ while maintaining the flatness of the flat band.
Focusing on the gap opening procedures, the positive (negative) $\delta$ opens the gap for the concave (convex) quadratic form of $t_1^\prime k_x^2 + t_2^\prime k_x k_y + t_3^\prime k_y^2$ as illustrated in Fig.~\ref{fig:gap_opening}(b).
Another important feature of the non-singular band touching is that any generic mass term $\mathcal{H}^\prime_{\v k} = m_x\sigma_x + m_y\sigma_y + m_z\sigma_z$ cannot make the flat band to have a nonzero Chern number after gap opening.
One can easily check that the Berry connection and curvature of $\mathcal{H}_{\v k}+\mathcal{H}^\prime_{\v k}$ are vanishing at all momenta.

\subsection{Singular band touching}

Unlike the case of the non-singular band touching whose low energy Hamiltonian can be described by a single Pauli matrix, the effective Hamiltonian for a singular band touching has at least two Pauli matrices and the flatness of the flat band is not guaranteed after degeneracy lifting.

Let us justify this statements by considering the general 2D continuum model around the singular touching point.
As analyzed in App.~\ref{app:low_energy} the general form of the quadratic band touching with a flat band in 2D can be described by
\begin{align}
\mathcal{H}_{\v k} =& (t_1 k_x^2 + t_2 k_x k_y + t_3 k_y^2)\sigma_z + (t_4k_xk_y +t_5k_y^2)\sigma_y \nonumber\\
&+t_6k_y^2\sigma_x + (b_1 k_x^2 + b_2 k_x k_y + b_3 k_y^2)\sigma_0,
\end{align}
which yields the singular touching only when $t_1$ and $t_4$ are nonzero due to the flatness condition $\mathrm{det}\tilde{\mathcal{H}}_{\v k} = 0$ as shown in App.~\ref{app:low_energy}.

After adding a perturbation $\mathcal{H}^\prime_{\v k}$ with three mass terms $m_{x,y,z}$, the flatness condition $\mathrm{det}(\mathcal{H}_{\v k}+\mathcal{H}_{\v k}^\prime) = 0$ yields four constraints on the masses given by (i) $b_1m_0 = t_1m_z$, (ii) $b_3 m_0 = m_z t_3 + m_y t_5 + m_x t_6$, (iii) $b_2 m_0 = m_z t_2 + m_y t_4$, and (iv) $m_0^2 = m_x^2 + m_y^2 + m_z^2$.
These constraints, together with the flatness condition $\mathrm{det}\tilde{\mathcal{H}}_{\v k} = 0$ of the unperturbed Hamiltonian, give us $t_4 = 0$ for both $t_1 = b_1$ and $t_1 =-b_1 $ when at least one of $m_i$'s is nonzero.
The final result $t_4 = 0$ contradicts the singular band touching condition, $t_4 \neq 0$, mentioned above, and this implies it is impossible to have a gap opening perturbation $\mathcal{H}^\prime_{\v k}$ that preserves the band flatness.
Namely, the singular flat band should become dispersive when the quadratic band touching is lifted by the generic mass term $\mathcal{H}^\prime_{\v k}$.
On the other hand, shifting the singular flat band to the opposite direction to have the band crossing with the quadratic band always leads to the splitting of the quadratic band touching into two Dirac points as shown in detail in App.~\ref{app:transition}.
Interestingly, such a deformation of the quadratic band crossing into two Dirac points is recently observed in the bosonic system on the honeycomb lattice made of polariton micropillars~\cite{Amo2018}.

Another interesting property of the singular touching distinguished from the non-singular one is that one can find a gap opening perturbation that would assign nonzero Chern number to the warped flat band.
For example, for the singular flat model described by
\begin{align}
\mathcal{H}_{\v k} = \frac{k_x^2 - k_y^2}{2}\sigma_z + k_xk_y\sigma_y + \frac{k_x^2 + k_y^2}{2}\sigma_0,
\end{align}
the Chern number of the warped flat becomes nonzero when the mass term $m\sigma_x$ is added. 
However, other kinds of mass term like $m\sigma_y$ and $m\sigma_z$ cannot open a gap but split the quadratic band crossing into two linear crossings.
Details for the calculation of the Chern number is in App.~\ref{sec:chern}.

One can see that the last conclusion holds also in the full lattice model by examining a perturbed kagome lattice model as an example.
The unperturbed Hamiltonian is described in (\ref{eq:kagome_ham}).
We add two kinds of mass terms $\mathcal{H}^{(1)} = \delta (\lambda_1+\lambda_6)$ and $\mathcal{H}^{(2)} = \delta (\lambda_2+\lambda_7)$ to (\ref{eq:kagome_ham}) where $\lambda_i$'s are the Gell-Mann matrices~\cite{GellMann} and $\delta$ is a real number.
While both perturbations lift the quadratic band crossing of the kagome lattice model, only 
the addition of $\mathcal{H}^{(2)}$ leads to the nearly flat band with a nonzero Chern number.

\section{Bulk-Boundary correspondence}\label{sec:bbc}

In this section, we show that the singular touching of the flat band has another crucial implication in the open boundary system.
To this end, we first demonstrate the existence of the non-compact states such as the non-contractible loop or planar states in the torus geometry when the flat band exhibits the singular touching.
Then we discuss how those non-contractible states are manifested as boundary modes when the system is terminated.
We confirm our correspondence by considering concrete examples.

\subsection{Non-contractible states in the bulk}

As discussed in Sec.~\ref{sec:irremovable_discon}, the flat band of the kagome lattice cannot be described completely by the CLSs, and the NLSs should be involved.
Naively, one may expect that the missing state can be complemented by adding two Bloch wave functions carrying the momentum at the singular point. Below we will show how to construct NLSs which are independent of the CLSs in both 2D and 3D cases. 

We first note that the Bloch wave function corresponding to the singular momentum does not contribute to the CLS as can be seen in (\ref{eq:cls_0}) since $\alpha_\mathbf{k} = 0$ at this momentum.
This means one can simply add the Bloch wave functions at the singular points to the incomplete set of CLSs to span the flat band completely.
Let us first consider the 2D flat band model with a singular point at $\mathbf{k} = (k_1^*,k_2^*)$.
For given $k_2=k_2^*$, we can perform a linear combination of Bloch wave functions with all possible $k_1$ including $k_1^*$.
While they are extended along $\mathbf{a}_2$ direction, an effective 1D Hamiltonian $\mathcal{H}_{k_1,k_2^*}$, considered as an 1D flat band model, ensures the existence of the linear combination of Bloch wave functions which is compact localized along $\mathbf{a}_1$ direction.
Thus the resulting wave function is a NLS extended along $\mathbf{a}_2$ direction. Similarly, one can also obtain another NLS extended along $\mathbf{a}_1$ direction.
These NLSs are linearly independent of the CLSs because they contain the Bloch wave function at $\mathbf{k} = (k_1^*,k_2^*)$ which is absent in CLSs.

In the case of a 3D flat band with an immovable discontinuity at $\v k = (k_x^*,k_y^*,k_z^*)$, one can perform a similar analysis by fixing two of $k_\alpha$'s at $k_\alpha^*$ $(\alpha=x,y,z)$.
In this case, the resultant wave function is extended along the two directions with the fixed momentum $k^*_\alpha$ whereas it is compact localized along the other direction.
We call such state a non-contractible planar state (NPS).

\begin{figure}
	\begin{center}
		\includegraphics[width=1\columnwidth]{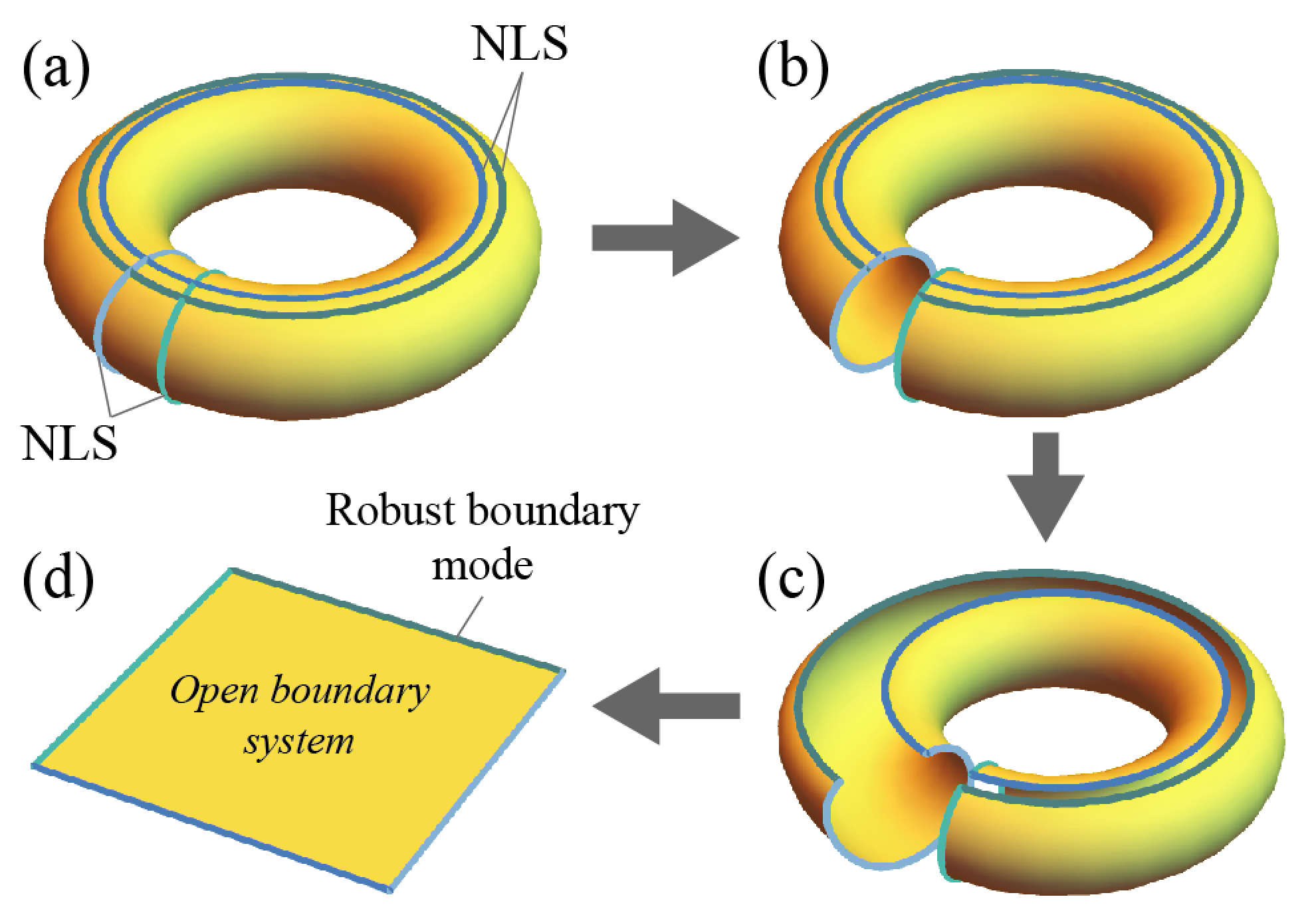}
	\end{center}
	\caption{ Schematic figures describing the robust boundary modes derived from the NLSs. (a) Four NLSs (blue lines) on a torus. (b,c) Deformation of the torus to a 2D plane with an open boundary by cutting the region between each pair of NLSs. (d) The original four NLSs become an eigenstate localized along the open boundary, which illustrates the bulk-boundary correspondence in singular flat bands.
	}
	\label{fig:torus}
\end{figure}

\subsection{Bulk-boundary correspondence}

The non-contractible states are realized on the surface of torus geometry reflecting the periodic boundary condition.
However, this geometry is hard to prepare experimentally.
So, we study the open boundary of the flat band model, and possible edge states.

One way of understanding the open boundary is to start from a torus geometry with a pair of the nearest neighboring NLSs along the poloidal direction and another nearest neighboring pair of NLSs along toroidal direction as illustrated in Fig.\ref{fig:torus}(a).
By cutting the regions between each pair of NLSs, the torus can be deformed to a 2D plane with open boundaries as shown in Fig.~\ref{fig:torus}(b-d).
Then, in the planar geometry, we obtain a boundary eigenmode with the same energy of the flat band.
So, the presence of the NLSs on the torus geometry guarantees the existence of the boundary mode in the planar geometry with the open boundary.

Another way of studying the open boundary is to exploit the incompleteness condition of the $N$ translated copies of CLSs on the torus geometry described by
\begin{align}
0 = \sum_{\v R} c_{\v R} | \chi_{\v R} \rangle,
\end{align}
where sum on $\v R$ is over all the lattice vectors in the system with the torus geometry.
In the finite system with an open boundary, on the other hand, this sum is not vanishing near the boundary, although it vanishes in the interior of the system far from the boundary.
The skin depth of $| \psi \rangle$ from the open boundary is usually less than the size of the CLS.
Since $| \chi_{\v R} \rangle$'s are all eigenstates, $| \psi \rangle$ is also an eigenmode localized around the open boundary of the system.

\begin{figure}
	\begin{center}
         \includegraphics[width=1\columnwidth]{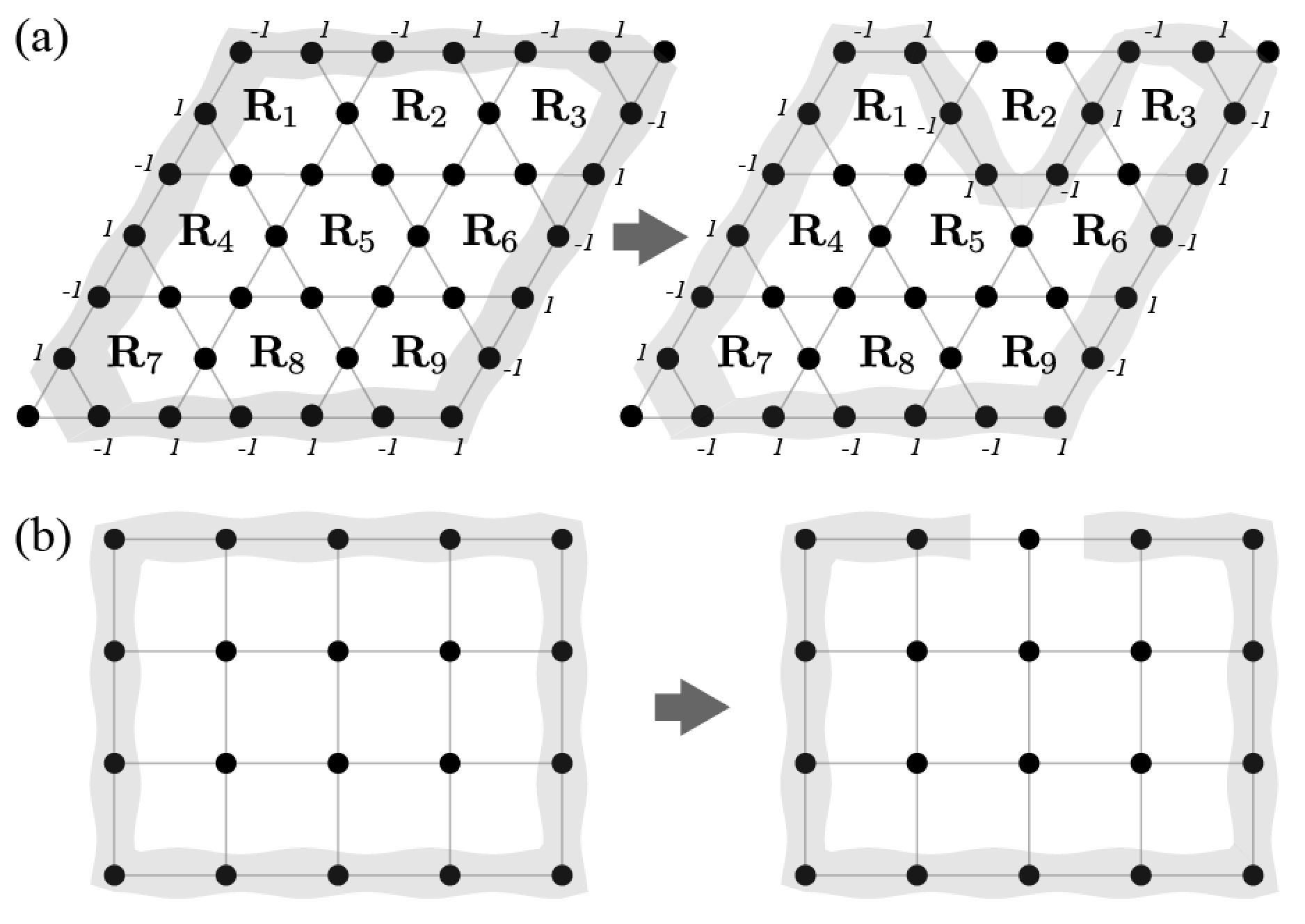}	\end{center}
	\caption{ (a) The robust boundary mode for a singular flat band in the kagome lattice. Adding a CLS merely deforms the shape of the boundary modes. (b) A fragile boundary mode for a non-singular flat band in the bilayer square lattice. The boundary mode can be disconnected by adding CLSs on the boundary.
	}
	\label{fig:robust_boundary}
\end{figure}

Thus obtained boundary state has some distinguishing properties compared with the usual topological boundary states.
First, the energy of this state is the same as that of the bulk flat band, which can be sharply contrasted to the usual in-gap boundary modes of conventional topological phases~\cite{Hatsugai1993,Mele2007,Fujimoto2012,Rhim2018,Rhim2016,Alvarez2018}.
As a result, this boundary mode cannot be observed by probing energy spectra.
Instead, one may examine the time-evolution of the system by generating the boundary mode as an initial state.
This might be possible in the bosonic systems such as the photonic crystal~\cite{Thomson2015,Amo2018, Chen2016a, Chen2016b} where we can prepare the initial input beam in the form of the boundary eigenmode, and then check its nondiffracting property.
Second, let us note that although the boundary state $|\psi\rangle$ is an eigenstate of the flat band model with an open boundary, it is not a new degree of freedom independent of CLSs.
In fact, $N$ translated copies of CLSs are all independent on the open geometry.
The distinct property of $|\psi\rangle$ of the singular flat band is that it cannot be disconnected by adding a finite number of CLSs additionally. 
Such a robustness of $|\psi\rangle$ against destructive interference originates from the fact that $|\psi\rangle$ is obtained by summing a macroscopic number of CLSs. 
On the other hand, the boundary mode of a non-singular flat band, which is merely a stack of CLSs along the boundary, can be easily disconnected by adding a few CLSs due to the destructive interference as shown in Fig.~\ref{fig:robust_boundary}(b).
This property originates from the non-contractible nature of the NLS or NRS.
In conclusion, the existence or absence this robust boundary mode is a crucial signature for distinguishing the singular or non-singular touching of the flat band.

\subsection{2D and 3D examples}

In this section, we introduce three concrete examples of flat band models.
Two of them have the singular touching of the flat band, while the other has a non-singular touching.
We show how those bulk properties are manifested as the non-contractible states or the robust boundary modes justifying our bulk-boundary correspondence.

As discussed in Sec.~\ref{sec:irremovable_discon}, the kagome lattice's flat band has a singular touching with the upper dispersive band.
As a result, the CLSs on the torus geometry cannot span the flat band completely, and two NLSs are complemented as shown in Fig.~\ref{fig:zigzag}(c).
In the planar geometry with the open boundary, as illustrated in Fig.~\ref{fig:robust_boundary}(a), we have a boundary eigenmode.
One can also check that this boundary state is actually constructed by the sum of all the possible translated copies of the CLS.

On the other hand, in the case of the bilayer square lattice studied in Sec.~\ref{sec:irremovable_discon}, which has a non-singular band touching, one cannot have the NLS or the robust boundary state.
Its CLS's amplitudes are nonzero only at the two sites of a vertical dimer.
Even if one can make a boundary mode by combining all the CLSs at the boundary, this can be disconnected by adding a CLS with opposite amplitudes as shown in Fig.~\ref{fig:robust_boundary}(d).
So, this is not the robust boundary state.

As a 3D flat band model with singular band touchings, let us consider a cubic lattice with three orbitals, denoted by $b_x,~b_y$, and $b_z$, per site described by
\begin{align}
H = \sum_{s=\pm 1}\sum_{\alpha,\beta,\gamma}\sum_{\mathbf{R}} \frac{s}{2} t_{\alpha\beta\gamma} b^\dag_{\alpha,\mathbf{R}+s\boldsymbol{\delta}_\gamma}b_{\beta,\mathbf{R}},
\end{align}
where $\alpha$, $\beta$, $\gamma$ run from $x$ to $z$, and $t_{xyz} = -t_{yxz} = 1$, $t_{zxy} = t_{xzy} = i$, and $t_{zyx} = t_{yzx} = -i$.
Here, $\boldsymbol{\delta}_\gamma = a\hat{\gamma}$.
The Fourier transformed Hamiltonian is then given by
\begin{align}
\mathcal{H}_\mathbf{k} = \bpm 0 & -i\sin k_z & \sin k_y \\ i\sin k_z & 0 & -\sin k_x \\ \sin k_y & -\sin k_x & 0 \epm,
\end{align}
whose eigenvector for the flat band is evaluated as
\begin{align}
\mathbf{v}_\mathbf{k} = \frac{1}{\alpha_\mathbf{k}}\bpm \sin k_x \\ \sin k_y \\ i\sin k_z \epm,
\end{align}
where 
\begin{align}
\alpha_\mathbf{k} = \sqrt{\sin^2k_x + \sin^2k_y + \sin^2k_z}.
\end{align}
This model shows a flat band at the zero energy and two dispersive bands.

\begin{figure}
	\begin{center}
		\includegraphics[width=1\columnwidth]{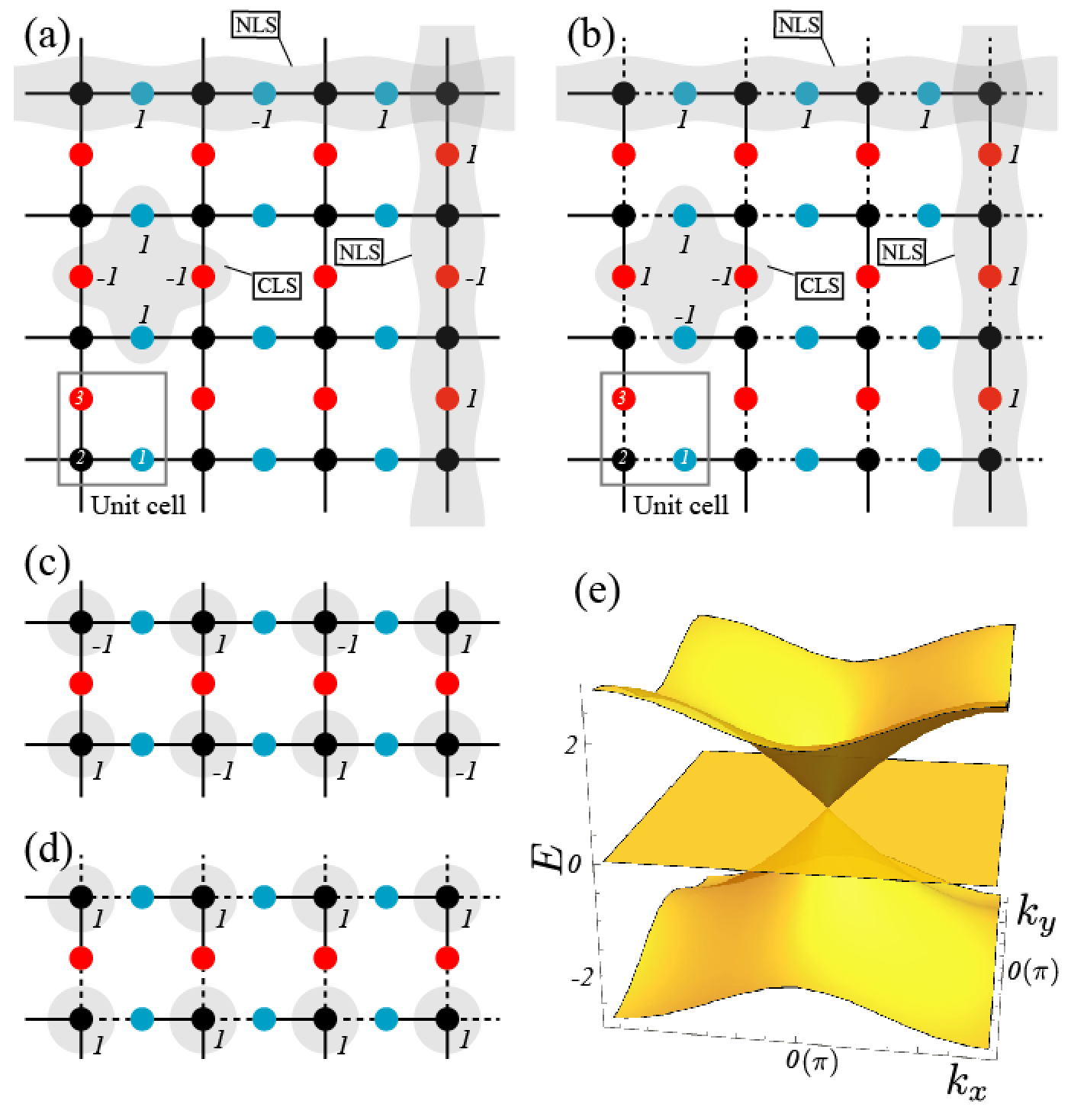}
	\end{center}
	\caption{ (a) The Lieb lattice. In the unit cell, we have three basis sites labeled by 1 (blue), 2 (black), and 3 (red). All the nearest hopping parameters corresponding to solid lines are $1$. The CLS and two NLSs are shown by the gray regions. (b) The modified Lieb lattice. The hopping amplitudes along the dashed and solid lines are $1$ and $-1$ respectively. In (c) and (d), we plot another extended state at the zero energy for the Lieb and the modified Lieb lattice model respectively. (e) The band structure of the Lieb and the modified Lieb lattice models which has a flat band at the zero energy. The Dirac point is located at $\v k = (0,0)$ in the modified Lieb lattice and at $\v k = (\pi,\pi)$ in the Lieb lattice.
	}
	\label{fig:lieb}
\end{figure}

The CLS around $\v R$ corresponding to $\mathbf{v}_\mathbf{k}$ is obtained as
\begin{align}
|\chi_{\mathbf{R}}\rangle \propto & i|x,\mathbf{R}+\boldsymbol{\delta}_x\rangle - i|x,\mathbf{R}-\boldsymbol{\delta}_x\rangle + i|y,\mathbf{R}+\boldsymbol{\delta}_y\rangle \nonumber \\
& - i|y,\mathbf{R}-\boldsymbol{\delta}_y\rangle - |z,\mathbf{R}+\boldsymbol{\delta}_z\rangle + |z,\mathbf{R}-\boldsymbol{\delta}_z\rangle.
\end{align}
As noted from the discontinuities of $\mathbf{v}_\mathbf{k}$ at $\mathbf{k} = (0,0,0)$, $(\pi,0,0)$, $(0,\pi,0)$, $(0,0,\pi)$, $(\pi,\pi,0)$, $(\pi,0,\pi)$, $(0,\pi,\pi)$, and $(\pi,\pi,\pi)$, $N=N_x N_y N_z$ number of translated copies of the CLS do not form a complete set.
Since only $\mathbf{k} =(0,0,0)$ is free from the even-odd effect of $N_\alpha$, let us consider, for simplicity, the case where $N_x$, $N_y$, and $N_z$ are all odd.
In this case, one can show that
\begin{align}
\sum_{\mathbf{R}} |\chi_{\mathbf{R}}\rangle = 0,
\end{align}
where the sum is over the whole lattice vectors on the 3-torus geometry of the cubic lattice.
Since only the Bloch wave function at $\mathbf{k}=0$ is missing when we construct the CLSs, the $N-1$ translated copies of the CLS are linearly independent and we need to find 3 complementary non-compact states to explain the $N+2$ degeneracy at the zero energy.
Note that we have triple degeneracy at $\v k = (0,0,0)$.

The three missing states are compensated by three NPSs with normal vectors $\hat{\alpha}=
\hat{x}$, $\hat{y}$, and $\hat{z}$ described by
\begin{align}
|\rho_\alpha \rangle = \sum_{\mathbf{R}\cdot\hat{\alpha} = 0} b^\dag_{\alpha,\mathbf{R}}|0\rangle,
\end{align}
where the sum is over all lattice vectors perpendicular to $\hat{\alpha}$.
When we consider a finite cube geometry, we have a robust boundary state which has finite amplitudes over all the six surfaces and vanishing amplitudes in the interior.

\subsection{About the geometric frustration}

When the NLSs are first discovered in the kagome lattice by Bergmann \textit{et al}, it was conjectured that the existence of NLSs might be closely related with the geometrical frustration of the hosting lattice~\cite{Balents2008}.
%
%
However, in our theory, the most fundamental origin of the NLSs is the Bloch wave function's discontinuity in momentum space.
We point out that although the geometric frustration could be helpful for realizing singular flat bands, it is not the generic origin of NLSs.
Let us clarify this point by constructing several model Hamiltonian explicitly as follows.

First, one can have singular flat band models on the lattices without geometrical frustration such as the Lieb lattice or the modified Lieb lattices described in App.~\ref{sec:lieb} and \ref{sec:modified_lieb}.
In the Lieb and the modified Lieb lattice models, the flat band has a singular touching at $\v k =(\pi,\pi)$ and $\v k =(0,0)$
This means that any $N$ number of CLSs are not linearly independent of each other, and some NLSs are required to be supplemented to span the flat band completely as shown in Fig.~\ref{fig:lieb}(a) and (b).
Unlike the kagome lattice model, the touching point is triply degenerate, which requires three additional states to  describe it.
In each case, we find two NLSs and one additional non-compact state which is completely extended occupying all the 2-sites (black sites) as illustrated in Fig.~\ref{fig:lieb}(c) and (d), relevant to the Lieb and modified Lieb lattices, respectively.

\begin{figure}
	\begin{center}
		\includegraphics[width=1\columnwidth]{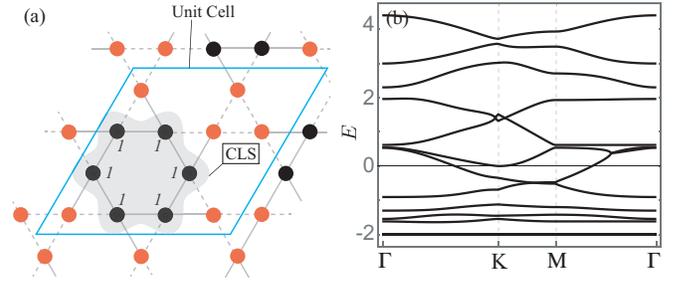}
	\end{center}
	\caption{ (a) kagome lattice with alternating hopping signs. The solid and dashed lines denote the hopping parameters -1 and 1 respectively. We use the red and black colors to distinguish sites with different on-site energies. The simplest CLS for this model is shown by the gray region. (b) The band structure when the on-site energies are zero for the black sites and 1 for the red sites.
	}
	\label{fig:kagome_2}
\end{figure}

As a second example, we construct a non-singular flat band model on a geometrically frustrated lattice, i.e., a modified kagome lattice model described in Fig.~\ref{fig:kagome_2}(a).
It has 12 basis sites in a unit cell, and contains two kinds of the nearest neighboring hopping processes with hopping amplitudes 1 and -1 as marked by the dashed and solid lines between neighboring sites.
The red and black sites have different onsite energies to each other.
As shown in the band structure in Fig.~\ref{fig:kagome_2}(b), this model has a flat band in the bottom which is completely separated from others without any band touching.
As discussed in the previous section, this kind of the flat band is a nonsingular type that can be spanned completely by $N$ translated copies of CLS without the help of NLSs despite the frustrated geometry.
The CLS is described in Fig.~\ref{fig:kagome_2}(a).

\begin{figure}
	\begin{center}
		\includegraphics[width=1\columnwidth]{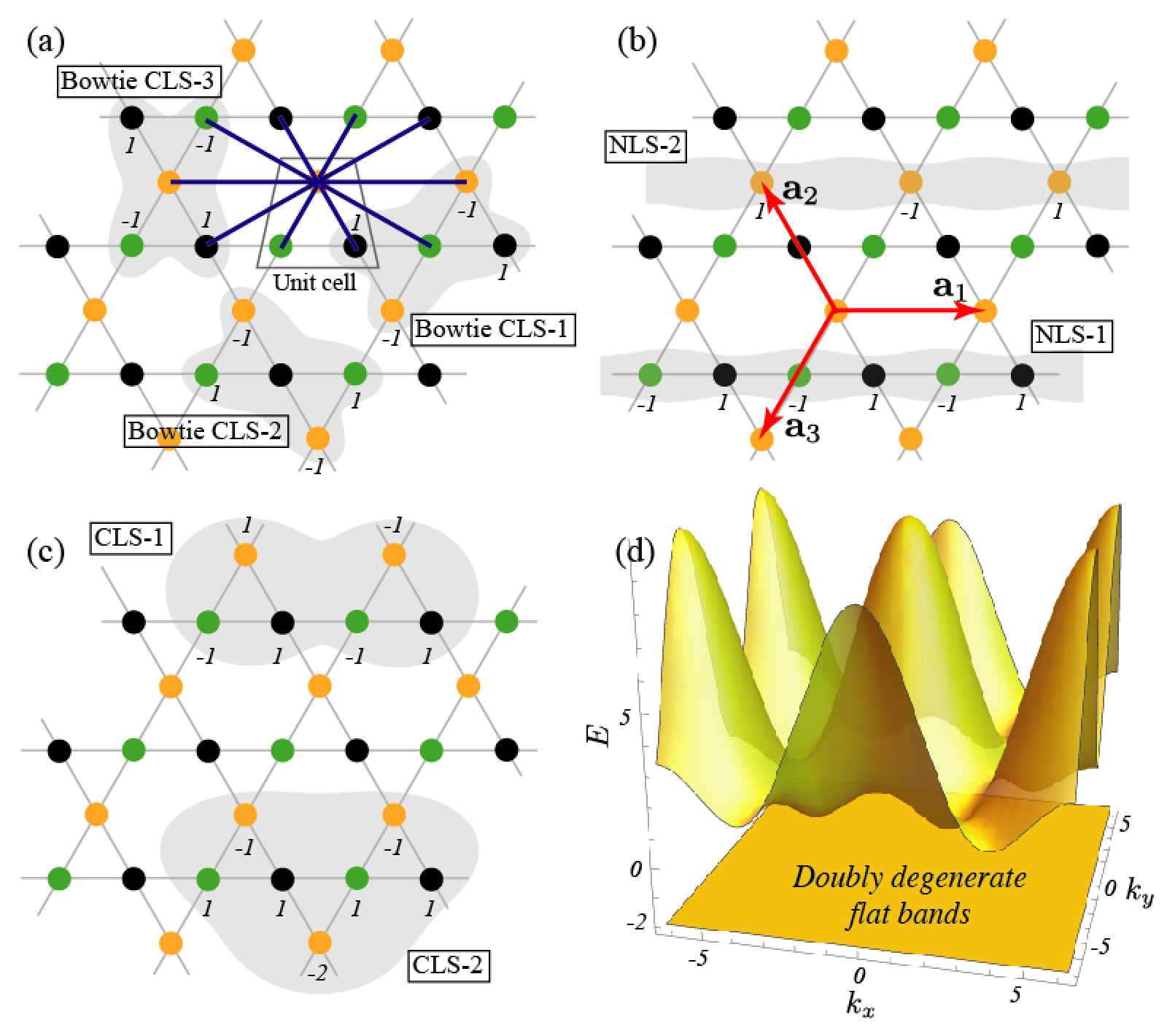}
	\end{center}
	\caption{ (a) The kagome-3 model is described. It has three basis sites labeled by 1 (yellow), 2 (green), and 3 (black). The hopping processes are allowed up to the third nearest  neighboring ones. We describe the hopping processes involved with the first site by thick blue lines. Three possible bowtie-shaped CLSs are drawn by the gray regions. The numbers in those regions are the amplitudes of CLSs. (b) The NLSs suggested by Bergman \textit{et al}. There are two more NLSs of the same types along $\v a_3$ direction. (c) Two new kinds of CLSs. The $N$ translated copies of each CLS form a complete set to span each flat band. (d) The band structure of the kagome-3 model. Two degenerate flat bands are separated from the dispersive one.
	}
	\label{fig:kagome_3}
\end{figure}

The kagome-3 model is another example with the frustrated geometry which has two completely degenerate flat bands separated from the dispersive one as shown in Fig.~\ref{fig:kagome_3}(d)~\cite{Balents2008}.
Refer to App.~\ref{app:kagome_3} for details.
In the original paper by Bergman \textit{et al}, they found two kinds of CLSs for this model as shown in Fig.~\ref{fig:kagome_3}(a), so called the bowtie CLS-1 and -3.
They noted that the $N$ translated copies of each of them do not constitute a complete set, and some NLSs are required to be supplemented.
They suggested four NLSs, two along $\v a_1$ and another two along $\v a_3$ directions, as two of them are depicted in Fig.~\ref{fig:kagome_3}(b).
At first glance, it sounds correct and consistent with our theory because the eigenvectors $\mathbf{v}^{(1)}_\mathbf{k}$ and $\mathbf{v}^{(2)}_\mathbf{k}$ for the two flat bands have singularities at $\mathbf{k} = (\pi,\pi/\sqrt{3})$ and $(\pi,-\pi/\sqrt{3})$ respectively.
However, we show in App.~\ref{app:kagome_3} that the NLSs actually can be constructed by the linear combinations of the bowtie CLSs by introducing another kind of the bowtie CLS denoted by bowtie CLS-2.
That is, the NLSs suggested by Bergman \textit{et al} can actually be disconnected by adding a finite number of bowtie CLSs as shown in Fig.~\ref{fig:kagome_3_nls}.
Let us note that Bergmann \textit{et al} have not considered bowtie CLS-2 as an independent state. This is because the sum of  six neighboring bowtie CLSs surrounding a hexagon, including each type of bowtie CLSs twice, vanishes. However, there is a caveat. Although two neighboring bowtie CLS-2s can be generated by the other four bowtie CLSs surrounding a hexagon, a single CLS-2 can still be independent of CLS-1 and 3.
%
%
Instead, we find another set of CLSs for the degenerate flat bands which are non-singular at all momenta by two linear combinations of $\mathbf{v}^{(1)}_\mathbf{k}$ and $\mathbf{v}^{(2)}_\mathbf{k}$ with momentum-dependent coefficients.
Since $\mathbf{v}^{(1)}_\mathbf{k}$ and $\mathbf{v}^{(2)}_\mathbf{k}$ have the singularities at different momenta, the singularities can be removed by this kind of momentum dependent mixing between them.
Two new CLSs (CLS-1 and CLS-2) are shown in Fig.~\ref{fig:kagome_3}(c).
This means that the flat bands of the kagome-3 model are non-singular type and we do not need any NLSs.
This example clearly shows that our approach based on the singularity of Bloch wave functions is more efficient and helpful to construct CLSs and NLSs as compared to the conventional approaches based on the intuition.

\section{General construction scheme for compact localized states}

When we consider the flat band model which can be treated analytically, the generic form of the eigenvector of the flat band is given by (\ref{eq:eigenvector}), and one can obtain the CLS by choosing $\alpha_\mathbf{k} = ( \sum_{q=1}^Q |w_\mathbf{k}|^2 )^{1/2}$.
However, if the analytic treatment is impossible and numerical studies are required, how to obtain such $\alpha_\mathbf{k}$?
If one could transform the Bloch basis into the $N$ translated CLSs, it has sometimes great advantages in studying the strongly correlated physics arising from the flat band model~\cite{Sarma2007}.

We show that 
\begin{align}
\alpha_\mathbf{k} = \frac{\mathrm{det}~\bar{\mathcal{H}}^{(p,p)}_\mathbf{k}}{v_{\mathbf{k},p}}\label{eq:alpha_gen}
\end{align}
can be a choice that makes all components of $\alpha_\mathbf{k}\mathbf{v}_\mathbf{k}$ the FSBP if there exists a component $v_{\mathbf{k},p}$ of $\v v_{\v k}$ such that the above formula is well-defined for all $\v k$.
Here, $\bar{\mathcal{H}}^{(p,p)}_\mathbf{k}$ is a $Q-1$ by $Q-1$ matrix obtained by eliminating the $p$-th row and column from $\bar{\mathcal{H}}_\mathbf{k}$.
To this end, we note that the following equation holds.
\begin{align}
\bar{\mathcal{H}}^{(p,p)}_\mathbf{k}\bpm \alpha_\mathbf{k}v_{\mathbf{k},1} \\ \vdots \\ \alpha_\mathbf{k}v_{\mathbf{k},Q} \epm^\prime = -\alpha_\mathbf{k} v_{\mathbf{k},p} \bpm \bar{\mathcal{H}}_\mathbf{k}|_{1,p} \\ \vdots \\ \bar{\mathcal{H}}_\mathbf{k}|_{Q,p} \epm^\prime,
\end{align}
which leads to
\begin{align}
\bpm \alpha_\mathbf{k}v_{\mathbf{k},1} \\ \vdots \\ \alpha_\mathbf{k}v_{\mathbf{k},Q} \epm^\prime = -\alpha_\mathbf{k} v_{\mathbf{k},p} \left(\bar{\mathcal{H}}^{(p,p)}_\mathbf{k}\right)^{-1} \bpm \bar{\mathcal{H}}_\mathbf{k}|_{1,p} \\ \vdots \\ \bar{\mathcal{H}}_\mathbf{k}|_{Q,p} \epm^\prime, \label{eq:other_components}
\end{align}
where the prime denotes the $p$-th component such as $\alpha_\mathbf{k}v_{\mathbf{k},p}$ and $\bar{\mathcal{H}}_\mathbf{k}|_{p,p}$ is excluded.
First, $\alpha_\mathbf{k}v_{\mathbf{k},p} = \mathrm{det}~\bar{\mathcal{H}}^{(p,p)}_\mathbf{k}$ is in the form of the FSBP.
According to the Cayley-Hamilton theorem, a general invertible matrix $A$ can be represented as 
\begin{align}
A^{-1} = \frac{-1}{\left| A \right|}\sum_{s=0}^{M-1}A^s\sum_{k_1,\cdots,k_{M-1}}\prod_{l=1}^{M-1} \frac{(-1)^{k_l}}{l^{k_l}k_l!}\mathrm{tr}\left(A^l\right)^{k_l},
\end{align}
where $|A| = \mathrm{det}~A$, and $k_l$ is all the nonnegative solutions of $s+\sum_{l=1}^{n-1}lk_l = n-1$ for each $s$~\cite{Kondratyuk}. 
This assures the other components in (\ref{eq:other_components}) are also in the form of the FSBP because the determinant factor $\mathrm{det}~\bar{\mathcal{H}}^{(p,p)}_\mathbf{k}$ in the denominator of the inverse matrix of $\bar{\mathcal{H}}^{(p,p)}_\mathbf{k}$ is cancelled by the same factor in $\alpha_\mathbf{k}$ in (\ref{eq:alpha_gen}).

As an example, let us consider the modified Lieb lattice of App.~\ref{sec:modified_lieb}.
From (\ref{eq:alpha_gen}) with $p=3$, we have $\alpha_{\v k} = E_+(\v k)^{1/2}(1-e^{ik_x})$ which makes $\alpha_{\v k}\v v_{\v k}$ in the form of the FSBP as follows.
\begin{align}
\alpha_{\v k}\v v_{\v k} = \bpm -1 + e^{ik_x}+e^{-ik_y}-e^{i(k_x-k_y)} \\ 0 \\ 2-e^{ik_x}-e^{-ik_x}\epm.
\end{align}
This leads to the CLS amplitude of the form
\begin{align}
\v A_{0,\v R} = \bpm -\delta^{\v R}_{(0,0)}+\delta^{\v R}_{(-1,0)} +\delta^{\v R}_{(0,1)} -\delta^{\v R}_{(-1,1)} \\ 0 \\ 2\delta^{\v R}_{(0,0)}-\delta^{\v R}_{(-1,0)} -\delta^{\v R}_{(1,0)} \epm.\label{eq:lieb_cls_2}
\end{align}
One can note that the size of the CLS described by (\ref{eq:lieb_cls_2}) is larger than (\ref{eq:lieb_cls_1}).
However, once we obtain a CLS of any size, one can easily get the smaller one by the linear combination between several translated copies of the CLS from (\ref{eq:alpha_gen}).
For instance, in this example, we have
\begin{align}
\v A_{0,\v R} + \v A_{(1,0),\v R} =& \bpm \delta^{\v R}_{(-1,0)}-\delta^{\v R}_{(-1,1)} -\delta^{\v R}_{(1,0)} +\delta^{\v R}_{(1,1)} \\ 0 \\ \delta^{\v R}_{(0,0)}-\delta^{\v R}_{(-1,0)} -\delta^{\v R}_{(2,0)} +\delta^{\v R}_{(1,0)} \epm \nonumber \\
=&\v A_{(-1,0),\v R}^{(0)} - \v A_{(1,0),\v R}^{(0)},
\end{align}
where $\v A_{\v R^\prime,\v R}^{(0)}$ is the smallest CLS defined in (\ref{eq:lieb_cls_1}).
Since $\v A_{(-1,0),\v R}^{(0)}$ and $\v A_{(1,0),\v R}^{(0)}$ are completely decoupled to each other, we can simply select one of them as a smaller CLS.

\section{General recipe to construct flat band models}

\subsection{Strategy}

In this section, we suggest a simple scheme to construct a flat band model with or without a singular touching in a controlled way.
A well-known method to construct a flat band tight binding model was to start from a nice miniarray of lattice sites which offers destructive interferences so that the CLS can be formed, and then build an infinite network of them with the translational symmetry~\cite{Vicencio2016}.
Although this scheme gives us an intuition about how the local structure of the lattice model specifically affects the destructive interferences of the wave function, one cannot determine whether the obtained model exhibits singular touching or not.

The overall strategy is as follows.
First, we prepare a lattice structure and a CLS as we want while the hopping parameters will be determined at the end.
At this stage we do not need to think about the normalization condition for the CLS, and we imagine that the CLS is in the form of $\alpha_\mathbf{k}\mathbf{v}_\mathbf{k}$ or (\ref{eq:fsbp}) as discussed in Sec.~\ref{sec:cls}.
The singular or non-singular nature of the flat band can be manipulated by making $\alpha_\mathbf{k}\mathbf{v}_\mathbf{k}$ vanishing or non-vanishig at a particular momentum $\v k=\v k^*$.
After we construct a complete set of eigenvectors including those of dispersive bands, we can easily build the relevant tight binding Hamiltonian.

\begin{figure}
	\begin{center}
		\includegraphics[width=1\columnwidth]{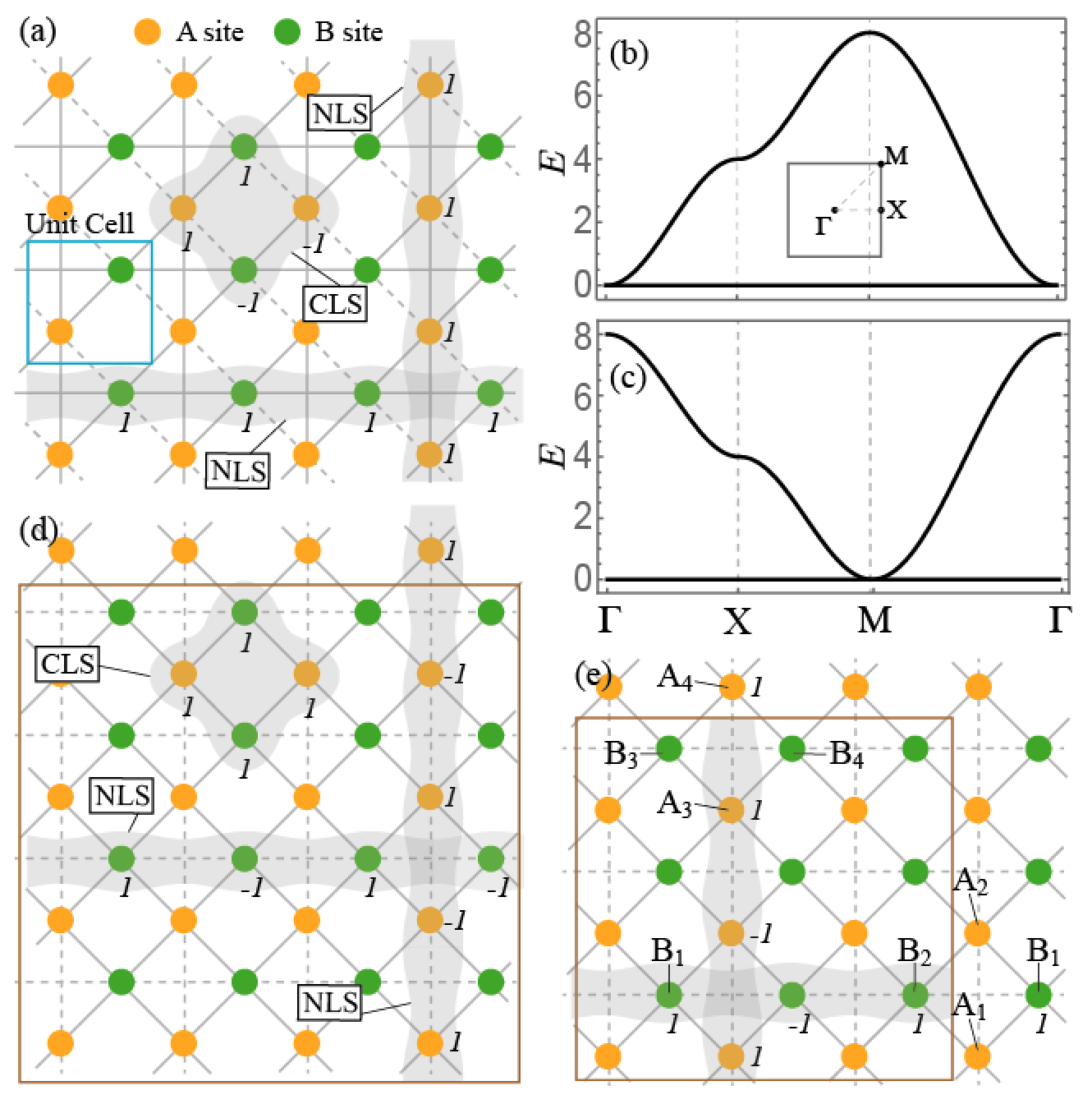}
	\end{center}
	\caption{ (a) The configuration of the checkerboard-I model. The dashed and solid lines represent the bonds with the hopping parameters $1$ and $-1$. The CLS and two NLSs are drawn by the gray regions with wave function's amplitudes on each site denoted by integer numbers. (b) and (c) correspond to band structures of the checkerboard model-I and -II respectively. (d) and (e) illustrate the configurations of the checkerboard-II model. In (d), we have even numbers of $N_x$ and $N_y$ in the whole system in the brown box. On the other hand, in (e) $N_x$ and $N_y$ are odd numbers. In (e), the NLSs in (d) are not eigenmodes anymore.
	}
	\label{fig:checker}
\end{figure}

\subsection{Singular touching at $\v k = \mathbf{(0,0)}$: checkerboard-I}

First, we construct a singular flat band model on the checkerboard lattice.
We design the flat band to have a singular touching at $\v k =0$.
While there can be numerous choices for CLSs, the simplest one can be obtained from
\begin{align}
\alpha_\mathbf{k}\mathbf{v}^{(1)}_\mathbf{k} = \bpm 1-e^{-ik_x} \\ 1-e^{ik_y}\epm,
\end{align}
where $\alpha_\mathbf{k} = (4 - 2\cos k_x -2\cos k_y)^{1/2}$ for $\mathbf{v}^{(1)}_\mathbf{k}$ to be normalized.
This has the amplitudes $1$ at both sites in the $\v R=0$ unit cell, and $-1$ at the A (B) site in the $\v R=(1,0)$ ($\v R=(0,-1)$) unit cell as shown in Fig.~\ref{fig:checker}(a).
The vanishing of $\alpha_\mathbf{k}\mathbf{v}^{(1)}_\mathbf{k}$ at $\v k =0$ implies that $\mathbf{v}^{(1)}_\mathbf{k}$ is discontinuous at there.
Another eigenvector orthogonal to $\mathbf{v}^{(1)}_\mathbf{k}$ can be obtained easily as
\begin{align}
\alpha_\mathbf{k}\mathbf{v}^{(2)}_\mathbf{k} = \bpm 1-e^{-ik_y} \\ -1+e^{ik_x} \epm,
\end{align}
which may correspond to another (dispersive) band.

Then, the Hamiltonian having $\mathbf{v}^{(1)}_\mathbf{k}$ and $\mathbf{v}^{(2)}_\mathbf{k}$ as eigenvectors can be composed as follows.
\begin{align}
\mathcal{H}_{\v k}\bpm v^{(1)}_{\v k,1} & v^{(2)}_{\v k,1} \\ v^{(1)}_{\v k,2} & v^{(2)}_{\v k,2}  \epm = \bpm 0 & E^{(2)}_{\v k} v^{(2)}_{\v k,1} \\ 0 & E^{(2)}_{\v k} v^{(2)}_{\v k,2}  \epm,
\end{align}
which leads to
\begin{align}
\mathcal{H}_{\v k} = \bpm 0 & E^{(2)}_{\v k} v^{(2)}_{\v k,1} \\ 0 & E^{(2)}_{\v k} v^{(2)}_{\v k,2}  \epm  \bpm v^{(1)}_{\v k,1} & v^{(2)}_{\v k,1} \\ v^{(1)}_{\v k,2} & v^{(2)}_{\v k,2}  \epm^\dag,
\end{align}
where we assume that the flat band is at the zero energy.
Here, $E^{(2)}_{\v k}$ is the energy dispersion of the other band.
To make the Hamiltonian in the form of the FSBP, $E^{(2)}_{\v k}$ should be chosen to be proportional to $\alpha_{\v k}^2$ as noted from the form of $\mathbf{v}^{(2)}_\mathbf{k}$ in the above.
In the simplest case where $E^{(2)}_{\v k} = \alpha_{\v k}^2$, the Hamiltonian becomes
\begin{align}
\mathcal{H}_\mathbf{k} = \bpm 2-2\cos k_y & -(1-e^{-ik_y})(1-e^{-ik_x}) \\ -(1-e^{ik_y})(1-e^{ik_x}) & 2-2\cos k_x \epm,
\end{align}
which has a flat band at the zero energy.
The relevant hopping amplitudes are shown in Fig.~\ref{fig:checker}(a).
One can check that the $N$ translated copies CLS are not independent of each other on the torus manifold because the sum of all CLSs vanishes as in the case of the kagome lattice.
The complementing noncontractible loop states are exhibited in Fig.~\ref{fig:checker}(a).

\begin{figure}
	\begin{center}
		\includegraphics[width=1\columnwidth]{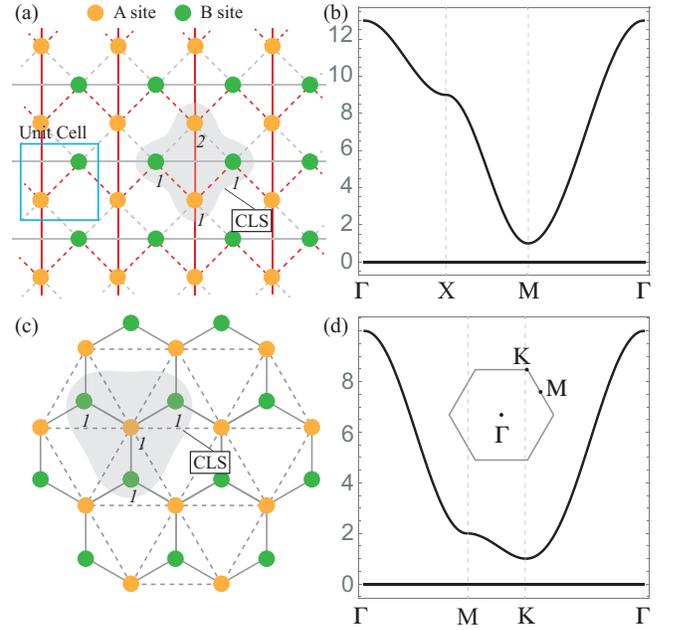}
	\end{center}
	\caption{ (a) The checkerboard-III model which hosts a non-singular flat band at the zero energy. A and B sites have different on-site energies 5 and 2, respectively. The solid red and gray lines represent the hopping parameters 2 and 1 while the dashes red and gray lines denote -2 and -1, respectively. Its band structure is drawn in (b). (c) The honeycomb lattice model yielding a non-singular flat band at the zero energy. A and B have the on-site energies 3 and 1, respectively. The solid and dashed lines mean the hopping parameters -1 and 1, respectively. Its band spectrum is shown in (d).
	}
	\label{fig:nonsingular_checker}
\end{figure}

\subsection{Singular touching at $\v k = (\boldsymbol\pi,\boldsymbol\pi)$: checkerboard-II}

On the same lattice, one can also make a Hamiltonian to have a singular flat band with the discontinuity at $\v k= (\pi,\pi)$.
To this end, we begin with
\begin{align}
\alpha_\mathbf{k}\mathbf{v}^{(1)}_\mathbf{k} = \bpm 1+e^{-ik_x} \\ 1+e^{ik_y}\epm,
\end{align}
where $\alpha_\mathbf{k} = (4 + 2\cos k_x + 2\cos k_y)^{1/2}$.
Another orthogonal eigenvector is found as
\begin{align}
\alpha_\mathbf{k}\mathbf{v}^{(2)}_\mathbf{k} = \bpm 1+e^{-ik_y} \\ -1-e^{ik_x} \epm.\label{eq:checker2_eigenvector}
\end{align}
With these, through the same procedure as before, we obtain the singular flat band Hamiltonian of the form
\begin{align}
\mathcal{H}_\mathbf{k} = \bpm 2+2\cos k_y & -(1+e^{-ik_y})(1+e^{-ik_x}) \\ -(1+e^{ik_y})(1+e^{ik_x}) & 2+2\cos k_x \epm.
\end{align}
The corresponding hopping amplitudes, the CLSs, and NLSs are described in Fig.~\ref{fig:checker}(d) or (e).
This model has a zero energy flat band touching with the dispersive upper band at $\mathbf{k} = (\pi,\pi)$ as plotted in Fig.~\ref{fig:checker}(c).
First, from (\ref{eq:checker2_eigenvector}), one can find the CLS $| \chi_\mathbf{R} \rangle$ with amplitudes $1$ at the $A$ sites in the unit cells at $\mathbf{R}$ and $\mathbf{R}+a\hat{x}$ and $B$ sites in the unit cells at $\mathbf{R}$ and $\mathbf{R}-a\hat{y}$ as shown in Fig.~\ref{fig:checker}(d).
However, $N$ translated copies of CLS do not form a complete set because $\alpha_\mathbf{k}^{(0)} = 0$ at $\mathbf{k} = (\pi,\pi)$ which is reflected by the multi-valuedness of the eigenvector at this momentum.
Indeed, one can show that
\begin{align}
0 = \sum_{\v R} (-1)^{\v R\cdot (\hat{x}+\hat{y})/a} |\chi_{\mathbf{R}}\rangle \quad (N_x,~N_y:\mathrm{even}), \label{eq:linear_dep}
\end{align}
where $\v R$ runs over the whole system indicated by the brown box in Fig.~\ref{fig:checker}(d) as an example, and $N_x$ and $N_y$ are the number of unit cells along $\hat{x}$ and $\hat{y}$ respectively.
The periodic boundary condition is applied to this system.
Two complementary NLSs extended along $x$ and $y$ directions are also depicted in Fig.~\ref{fig:checker}(d).
However, if at least one of $N_x$ or $N_y$ is odd, the $N$ translated CLSs form a complete set.
This is consistent with the fact that $\alpha_\mathbf{k}^{(0)} \neq 0$ with $k_x = 2\pi n_x/N_x$ and $k_y = 2\pi n_y/N_x$ where $n_x$ and $n_y$ are integers.

\subsection{Non-singular band touching: checkerboard-III}

In the same lattice, we can also have a flat band completely separated from other dispersive bands by assigning proper hopping parameters.
We start from the CLS without singularities given by
\begin{align}
\alpha_\mathbf{k}\mathbf{v}^{(1)}_\mathbf{k} = \bpm 1+e^{-ik_x} \\ 2+e^{ik_y}\epm,
\end{align}
where $\alpha_\mathbf{k} = (7 + 2\cos k_x + 4\cos k_y)^{1/2}$.
Then, another eigenvector orthogonal to the above is found to be
\begin{align}
\alpha_\mathbf{k}\mathbf{v}^{(2)}_\mathbf{k} = \bpm 2+e^{-ik_y} \\ -1-e^{ik_x}\epm.
\end{align}
Repeating the same process as before, the Hamiltonian having $\mathbf{v}^{(1)}_\mathbf{k}$ and $\mathbf{v}^{(2)}_\mathbf{k}$ as the eigenvectors, and the flat band at the zero energy is obtained as
\begin{align}
\mathcal{H}_\mathbf{k} = \bpm 5+4\cos k_y & -(1+e^{-ik_x})(2+e^{-ik_y}) \\ -(1+e^{ik_x})(2+e^{ik_y}) & 2+2\cos k_x \epm.
\end{align}
The configuration of the hopping processes relevant to this Hamiltonian is illustrated in Fig.~\ref{fig:nonsingular_checker}(a).
As shown in Fig.~\ref{fig:nonsingular_checker}(b), the flat band at the zero energy is completely separated from another dispersive band which implies that the flat band is non-singular.
Also the $N$ translated copies of the CLS represented by the gray region in Fig.~\ref{fig:nonsingular_checker}(a) form a complete set spanning the flat band.

\subsection{Non-singular band touching: honeycomb}

In the section, we consider a different lattice structure, the honeycomb lattice.
We construct a Hamiltonian possessing a flat band which is completely separated from another.
We consider a CLS corresponding to
\begin{align}
\alpha_\mathbf{k}\mathbf{v}^{(1)}_\mathbf{k} = \bpm 1 \\ 1+e^{i\v a_1\cdot\v k}+e^{i(\v a_1-\v a_2)\cdot\v k}\epm,
\end{align}
where $\v a_1 = (1/2,\sqrt{3}/2)$, $\v a_1 = (-1/2,\sqrt{3}/2)$, and $\alpha_{\v k} = (4 + 2\cos k_x + 2\cos k_x/2 \cos \sqrt{3}k_y/2)^{1/2}$.
The shape of the CLS is drawn in Fig.~\ref{fig:nonsingular_checker}(c).
Since it has no zeros in momentum space, $\mathbf{v}^{(1)}_\mathbf{k}$ is non-singular.
Another orthogonal eigenvector is given by
\begin{align}
\alpha_\mathbf{k}\mathbf{v}^{(2)}_\mathbf{k} = \bpm  1+e^{-i\v a_1\cdot\v k}+e^{-i(\v a_1-\v a_2)\cdot\v k} \\ -1 \epm.
\end{align}
Then, repeating the same procedure, we obtain
\begin{align}
\mathcal{H}_{\v k}|_{1,1} =& 3+2\cos k_x+4\cos\frac{k_x}{2}\cos\frac{\sqrt{3}k_y}{2}, \\
\mathcal{H}_{\v k}|_{1,2} =& \mathcal{H}_{\v k}^*|_{2,1} = -1-e^{-ik_x}-e^{-i\frac{1}{2}(k_x+\sqrt{3}k_y)}, \\
\mathcal{H}_{\v k}|_{2,2} =& 1,
\end{align}
whose hopping processes in real space are shown in Fig.~\ref{fig:nonsingular_checker}(c).
Note that the next nearest neighbor hopping is allowed only between A sites.
As plotted in Fig.~\ref{fig:nonsingular_checker}(d), the flat band is completely separated from another dispersive band as expected from the non-singular property of the Bloch eigenfunction of the flat band.

\subsection{General scheme}

Up to now, we have focused on simple 2 by 2 Hamiltonian matrices to demonstrate the general recipe to construct flat band models. 
However, the recipe can be generally applied to any Hamiltonian with an arbitrary size.
We first design a CLS for a flat band by writing down a unnormalized eigenvector $\alpha_{\v k}\v v^{(0)}$ of size $Q$ (the number of orbitals in a unit cell).
At this stage, we already determine whether the flat band is singular or not.
If the flat band is singular, $\alpha_{\v k}\v v^{(0)}$ vanishes at a momentum while if it is non-singular, $\alpha_{\v k}\v v^{(0)}$ is nonzero for all momenta.
Then, we should find $Q-1$ other eigenvectors orthonormal to each other as well as to $\v v^{(0)}$, denoted by $\mathbf{v}_\mathbf{k}^{(q)}$ ($1\leq q \leq Q-1$), to construct a full tight binding Hamiltonian.
There are arbitrarily many choices for such set of eigenvectors, and we obtain different tight binding models depending on the choice.
An option is to apply the Gram-Schmidt process to obtain the set of orthonormal wave functions from any set of linearly independent vectors such as $(1,0,\cdots,0)$, $(0,1,\cdots,0)$, $\cdots$, and $\mathbf{v}_\mathbf{k}^{\mathrm{(0)}}$.
Then, our target Hamiltonian satisfies
\begin{align}
\mathcal{H}_\mathbf{k} \mathcal{U}_\mathbf{k} = \mathcal{V}_\mathbf{k}, \label{eq:hamiltonian}
\end{align}
where
\begin{align}
\mathcal{U}_\mathbf{k} = \bpm v_{\mathbf{k},1}^{\mathrm{CLS}} & v^{(1)}_{\mathbf{k},1} & \cdots & v^{(Q-1)}_{\mathbf{k},1} \\  v_{\mathbf{k},2}^{\mathrm{CLS}} & v^{(1)}_{\mathbf{k},2} & \cdots & v^{(Q-1)}_{\mathbf{k},2} \\ \vdots & \vdots & \ddots & \vdots \\ v_{\mathbf{k},Q}^{\mathrm{CLS}} & v^{(1)}_{\mathbf{k},Q} & \cdots & v^{(Q-1)}_{\mathbf{k},Q}  \epm,
\end{align}
which is composed of the CLS in the first column and other orthonormal vectors in other columns, and
\begin{align}
\mathcal{V}_\mathbf{k} = \bpm 0 & E_\mathbf{k}^{(1)} v^{(1)}_{\mathbf{k},1} & \cdots & E_\mathbf{k}^{(Q-1)} v^{(Q-1)}_{\mathbf{k},1} \\  0 & E_\mathbf{k}^{(1)} v^{(1)}_{\mathbf{k},2} & \cdots & E_\mathbf{k}^{(Q-1)} v^{(Q-1)}_{\mathbf{k},2} \\ \vdots & \vdots & \ddots & \vdots \\ 0 & E_\mathbf{k}^{(1)} v^{(1)}_{\mathbf{k},Q} & \cdots & E_\mathbf{k}^{(Q-1)} v^{(Q-1)}_{\mathbf{k},Q}  \epm,
\end{align}
where $E^{(q)}_\mathbf{k}$ is the possible eigenenergy of $\mathbf{v}_\mathbf{k}^{(q)}$.
By multiplying $\mathcal{U}_\mathbf{k}^{-1} = \mathcal{U}_\mathbf{k}^\dag$ to both sides of (\ref{eq:hamiltonian}), we have
\begin{align}
\mathcal{H}_\mathbf{k}\big|_{ij} = \sum_{q=1}^{Q-1} E^{(q)}_\mathbf{k} v^{(q)}_{\mathbf{k},i} v^{(q)*}_{\mathbf{k},j}.\label{eq:obtained_hamiltonian}
\end{align}
While $E^{(q)}_\mathbf{k}$ also can be chosen freely, it is required to make all the elements of the Hamiltonian in the form of the FSBP.
In general, not all the choices of $\mathbf{v}_\mathbf{k}^{(q)}$ allow us to have such $E^{(q)}_\mathbf{k}$.
However, if $\mathbf{v}_\mathbf{k}^{(q)}$ is obtained from the Gram-Schmidt process starting from the initial basis vectors in the form of the FSBP, such $E^{(q)}_\mathbf{k}$ exists.
This is because thus obtained $\mathbf{v}_\mathbf{k}^{(q)}$'s are in the form of (\ref{eq:eigenvector}).
This can be shown by the mathematical induction as follows.
Let us denote the initial unnormalized vectors as $\mathbf{u}_\mathbf{k}^{(q)}$ whose components are in the form of the FSBP.
Then, we have
\begin{align}
\mathbf{v}_\mathbf{k}^{(1)} \propto \mathbf{u}_\mathbf{k}^{(1)} - \left[ \left(\mathbf{v}_\mathbf{k}^{\mathrm{(0)}}\right)^\dag\cdot \mathbf{u}_\mathbf{k}^{(1)} \right] \mathbf{v}_\mathbf{k}^{\mathrm{(0)}},
\end{align}
which can be transformed to the form of the FSBP by multiplying the factor $(\alpha_\mathbf{k}^\mathrm{(0)})^2$.
This leads to the expression
\begin{align}
\mathbf{v}_\mathbf{k}^{(1)} = \frac{1}{\sqrt{\sum_{q=1}^Q |w_{\mathbf{k},q}^{(1)}|^2}}\bpm w_{\mathbf{k},1}^{(1)} \\ \vdots \\  w_{\mathbf{k},Q}^{(1)} \epm = \frac{1}{\alpha_\mathbf{k}^{(1)}}\bpm w_{\mathbf{k},1}^{(1)} \\ \vdots \\  w_{\mathbf{k},Q}^{(1)} \epm,
\end{align}
where $w_{\mathbf{k},q}^{(1)}$ is in the form of the FSBP.
Let us assume that $\mathbf{v}_\mathbf{k}^{(q)}$ is also represented as
\begin{align}
\mathbf{v}_\mathbf{k}^{(q)} = \frac{1}{\sqrt{\sum_{p=1}^Q |w_{\mathbf{k},p}^{(q)}|^2}}\bpm w_{\mathbf{k},1}^{(q)} \\ \vdots \\  w_{\mathbf{k},Q}^{(q)} \epm = \frac{1}{\alpha_\mathbf{k}^{(1)}}\bpm w_{\mathbf{k},1}^{(q)} \\ \vdots \\  w_{\mathbf{k},Q}^{(q)} \epm.
\end{align}
Then, $\mathbf{v}_\mathbf{k}^{(q+1)}$ is obtained as
\begin{align}
\mathbf{v}_\mathbf{k}^{(q+1)} \propto \mathbf{u}_\mathbf{k}^{(q+1)} - \sum_{p=0}^q \left[ \left(\mathbf{v}_\mathbf{k}^{\mathrm{(p)}}\right)^\dag\cdot \mathbf{u}_\mathbf{k}^{(q+1)} \right] \mathbf{v}_\mathbf{k}^{\mathrm{(p)}}.
\end{align}
Multiplying the factor $\prod_{p=0}^{q} (\alpha_\mathbf{k}^{(p)})^2$, one can have $\mathbf{v}_\mathbf{k}^{(q+1)}$ in the form of the FSBP, and it can be written in the same form of (\ref{eq:eigenvector}) with the normalization coefficient.
Once the $Q$ orthonormal basis vectors are prepared, the eigenenergies are simply of the form
\begin{align}
E^{(q)}_\mathbf{k} = F_\mathbf{k} \times (\alpha_\mathbf{k}^{(q)})^2,
\end{align}
where $F_\mathbf{k}$ is an arbitrary function of $\mathbf{k}$ in the form of the FSBP.
This makes $\mathcal{H}_\mathbf{k}\big|_{ij}$ also the FSBP form because $v^{(q)}_{\mathbf{k},i}$ and $v^{(q)*}_{\mathbf{k},j}$ in (\ref{eq:obtained_hamiltonian}) share the same $\alpha^{(q)}_\mathbf{k}$ factor.

\section{Conclusions}

In this work, we suggest a completely different approach for analyzing the flat bands by focusing on the singularity of Bloch wave functions in momentum space, which is alternative to the conventional approach based on the local symmetries of the lattice model in real space.
Our scheme offers a unified way to analyze the flat band models regardless of their dimensionality, detailed lattice structures, and symmetries.
We show that the existence or absence of the immovable discontinuities of the Bloch wave function in momentum space, which is generated by the band touching, determines the singular or non-singular character of the flat band.
In the case of a non-singular touching, one can always find a mass term as a perturbation that lifts the degeneracy while keeping the band flatness. 
On the other hand, the singular touching is protected by the band flatness.
If the degeneracy at the band crossing is lifted, the resultant nearly flat band can gain a nonzero Chern number.
One can construct a complete set of CLSs for a non-singular flat band whereas, in the case of singular flat bands, the $N$ translated copies of CLSs are incomplete due to the singularity of the Bloch wave functions.

Furthermore, we demonstrate that the presence of the discontinuity of the Bloch wave function implies that we have the robust boundary mode as an eigenstate at the open boundary of the system.
Interestingly, this mode has the same energy as the flat band, not located in the gap.
We suggest that this new kind of the bulk-boundary correspondence can be observed experimentally in the bosonic systems like the photonic crystals~\cite{Thomson2015,Amo2018, Chen2016a, Chen2016b} where it recently has become possible to observe the CLSs and flat bands.
In this system, one might also prepare the robust boundary state as an initial state, and then observe its evolution in time to confirm the compact localization of the eigenmode.

Our finding demonstrates a new perspective on the role of the bulk Bloch wave functions to characterize flat band systems.
Although flat bands are generally expected to be topologically trivial, their Bloch wave function still contains the key information about the singular nature of the associated band crossing.
Finally, we note that our theory naturally leads to systematic schemes useful for the construction of the CLSs and flat band tight binding models.
Up to now, flat band models have been constructed based on physical intuition, which cannot be generally applied to complicated systems with long range hopping, in high dimensions, or not tractable analytically.
However, our schemes overcome those difficulties so that one might have advantages in designing the flat band models or finding compact localized basis for the study of strongly interacting systems.

\acknowledgements
J.-W.R was supported by IBS-R009-D1.
B.-J.Y. was supported by the Institute for Basic Science in Korea (Grant No. IBS-R009-D1) and Basic Science Research Program through the National Research Foundation of Korea (NRF) (Grant No. 0426-20170012, No.0426-20180011), the POSCO Science Fellowship of POSCO TJ Park Foundation (No.0426-20180002), and the U.S. Army Research Office under Grant Number W911NF-18-1-0137.

\appendix

\newpage

\section{Existence of the compact localized state}\label{app:cls_existence}

In this section, we show that a choice of $\alpha_\mathbf{k}$, which makes $|\chi_\mathbf{R}\rangle$ compact localized, always exists if the band is flat over the whole Brillouin zone.
It is equivalent to prove that the kernel of the matrix $\bar{\mathcal{H}}_\mathbf{k}=\mathcal{H}_\mathbf{k}-\epsilon_0\mathcal{I}$, where $\epsilon_0$ is the energy of the flat band, has an element $\mathbf{x}_\mathbf{k}$ in the form of the FSBP.
$\alpha_\mathbf{k}$ can be obtained from $\mathbf{x}_\mathbf{k} = \alpha_\mathbf{k}\mathbf{v}_\mathbf{k}$.
We assume that the hopping range of the tight binding model is finite, so that all elements of $\mathcal{H}_\mathbf{k}$ are in the form of the FSBP.

This can be shown by the mathematical induction.
First, when $\bar{\mathcal{H}}_\mathbf{k}$ is a $2\times 2$ nonzero matrix, one can find a solution of the form
\begin{align}
\mathbf{x}_\mathbf{k} = \bpm \bar{\mathcal{H}}_\mathbf{k}|_{1,2} \\ -\bar{\mathcal{H}}_\mathbf{k}|_{1,1} \epm,
\end{align}
where its elements are in the form of the FSBP.
One can easily see that the first row of $\bar{\mathcal{H}}_\mathbf{k}\mathbf{x}_\mathbf{k}$ is vanishing. 
Then the second row of $\bar{\mathcal{H}}_\mathbf{k}\mathbf{x}_\mathbf{k}$ is also zero because we assume that there exists a flat band at $\epsilon_0$ which guarantees $\mathrm{det}\bar{\mathcal{H}}_\mathbf{k} = 0$.

Then, let us assume that the kernel of any $(Q-1)\times (Q-1)$ matrix $\bar{\mathcal{H}}_\mathbf{k}$ always possesses a vector $\mathbf{x}_\mathbf{k}$ in the form of the FSBP.
Then, we consider a $Q\times Q$ matrix $\bar{\mathcal{H}}_\mathbf{k}$ with a flat band at $\epsilon_0$.
We assume that every column of $\bar{\mathcal{H}}_\mathbf{k}$ contains at least one nonzero element.
Otherwise, the problem just reduces to the one considering $(Q-1)\times (Q-1)$ matrix which is already assumed to have a solution in the form of the FSBP.
In the system of $Q$ homogeneous equations given by $\bar{\mathcal{H}}_\mathbf{k}\mathbf{x}_\mathbf{k}=0$, one can eliminate $x_{\mathbf{k},Q}$, the $Q$-th component of $\mathbf{x}_\mathbf{k}$, by multiplying $\bar{\mathcal{H}}_\mathbf{k}|_{1,Q}$ and $\bar{\mathcal{H}}_\mathbf{k}|_{q,Q}$ to the $q$-th and the first rows of $\bar{\mathcal{H}}_\mathbf{k}\mathbf{x}_\mathbf{k}=0$ and subtracting between them for all $q$'s between $2$ and $Q$.
Then, we obtain a system of $(Q-1)$ number of equations for $x_{\mathbf{k},1},~x_{\mathbf{k},2},\cdots,x_{\mathbf{k},Q-1}$ given by
\begin{align}
0=\sum_{j=1}^{Q-1} \left( \bar{\mathcal{H}}_\mathbf{k}|_{q,j}\bar{\mathcal{H}}_\mathbf{k}|_{1,Q} -  \bar{\mathcal{H}}_\mathbf{k}|_{1,j}\bar{\mathcal{H}}_\mathbf{k}|_{q,Q} \right) x_{\mathbf{k},j}.\label{eq:gaussian}
\end{align}
One can note that the above can be represented by $\bar{\mathcal{K}}_\mathbf{k}\mathbf{x}_\mathbf{k}=0$, where $\bar{\mathcal{K}}_\mathbf{k}|_{q,j}=\bar{\mathcal{H}}_\mathbf{k}|_{q,j}\bar{\mathcal{H}}_\mathbf{k}|_{1,Q} -  \bar{\mathcal{H}}_\mathbf{k}|_{1,j}\bar{\mathcal{H}}_\mathbf{k}|_{q,Q}$.
$\bar{\mathcal{K}}_\mathbf{k}$ is a $(Q-1)\times (Q-1)$ matrix with elements in the form of the FSBP.
This implies that $x_{\mathbf{k},1},~x_{\mathbf{k},2},\cdots,x_{\mathbf{k},Q-1}$ can be chosen to be in the form of the FSBP.
If we denote such solution for (\ref{eq:gaussian}) as $x_{\mathbf{k},j}=y_{\mathbf{k},j}$ and assume that the nonzero component of the $Q$-th column of $\bar{\mathcal{H}}_\mathbf{k}$ is $\bar{\mathcal{H}}_\mathbf{k}|_{p,Q}$, $x_{\mathbf{k},j} = y_{\mathbf{k},j} \bar{\mathcal{H}}_\mathbf{k}|_{p,Q} $ is also a nontrivial solution of (\ref{eq:gaussian}) in the form of the FSBP.
Finally, the remaining last component of $\mathbf{x}_\mathbf{k}$ is determined from the $p$-th row of the equation $\bar{\mathcal{H}}_\mathbf{k}\mathbf{x}_\mathbf{k}=0$ as follows.
\begin{align}
x_{\mathbf{k},Q} = -\sum_{j=1}^{Q-1} \bar{\mathcal{H}}_\mathbf{k}|_{p,j} y_{\mathbf{k},j}.
\end{align}
This is also in the form of the FSBP.
Again, from the relation $\mathbf{x}_\mathbf{k} = \alpha_\mathbf{k}\mathbf{v}_\mathbf{k}$, we find the multiplying factor $4\alpha_\mathbf{k}$ that makes the eigenvector in the form of the FSBP.
In the proof, the key point is that all matrix elements of $\bar{\mathcal{H}}_\mathbf{k}=\mathcal{H}_\mathbf{k}-\epsilon_{n,\mathbf{k}}\mathcal{I}$ are in the form of the FSBP if the band $\epsilon_{n,\mathbf{k}}$ is flat. 
This allows us to have a solution of $\bar{\mathcal{H}}_\mathbf{k}\mathbf{x}_\mathbf{k}=0$ in the form of the FSBP.

\begin{figure}
	\begin{center}
		\includegraphics[width=1\columnwidth]{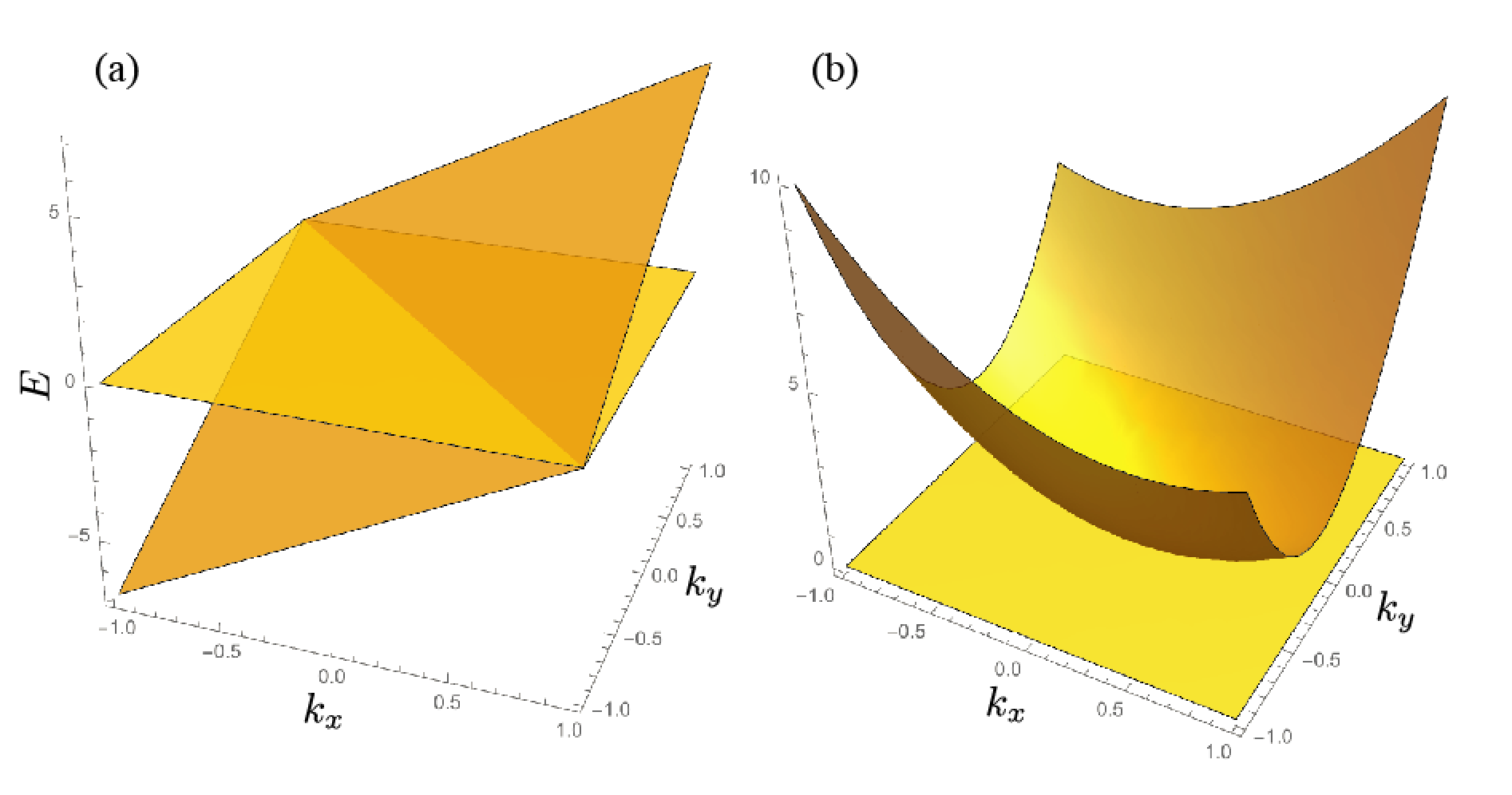}
	\end{center}
	\caption{ (a) The band structure of the Hamiltonian (\ref{eq:non_singular_eff_ham}) with $\alpha_i = c_y = c_z =1$, and $c_0 = \sqrt{3}$. (b) The band spectrum of the Hamiltonian (\ref{eq:singular_eff_ham}) with $t_1 = t_2 = t_3 = t_5 = b_1 = b_2 = 1$, $t_4 = 2$, $t_6 = \sqrt{7}$, and $b_3 = 3$.
	}
	\label{fig:eff_bands}
\end{figure}

\section{Low energy description of the flat band touching}\label{app:low_energy}

\subsection{Linear touching}
We show that if a dispersive band is touching with a flat band linearly, the flat band must be non-singular.
Since all 1D flat bands are non-singular (Sec.~\ref{sec:irremovable_discon}), we begin with a general two dimensional $2\times 2$ effective Hamiltonian of the form
\begin{align}
\mathcal{H}_{\v k} =& (b_xk_x+b_yk_y)\sigma_0 + (\alpha_xk_x+\alpha_yk_y)\sigma_x \nonumber \\
&+(\beta_xk_x+\beta_yk_y)\sigma_y +(\gamma_xk_x+\gamma_yk_y)\sigma_z,
\end{align}
around the band touching point.
The condition for a flat band is at the zero energy is described by an equation 
\begin{align}
\mathrm{det}\mathcal{H}_{\v k} = 0.
\end{align}
This leads to
\begin{align}
b_x^2 =& \alpha_x^2 + \beta_x^2 +\gamma_x^2, \\
b_y^2 =& \alpha_y^2 + \beta_y^2 +\gamma_y^2, \\
b_xb_y =& \alpha_x\alpha_y + \beta_x\beta_y + \gamma_x\gamma_y,
\end{align}
which lead to
\begin{align}
0 =& (\alpha_x\beta_y-\beta_x\alpha_y)^2 + (\alpha_x\gamma_y-\gamma_x\alpha_y)^2 \nonumber\\
&+ (\beta_x\gamma_y-\gamma_x\beta_y)^2.
\end{align}
If we define $\boldsymbol\alpha = (\alpha_x,\alpha_y,0)$, $\boldsymbol\beta = (\beta_x,\beta_y,0)$, and $\boldsymbol\gamma = (\gamma_x,\gamma_y,0)$, we obtain
\begin{align}
\boldsymbol\alpha\times\boldsymbol\beta = \boldsymbol\alpha\times\boldsymbol\gamma = \boldsymbol\gamma\times\boldsymbol\beta = 0,
\end{align}
and we conclude that
\begin{align}
\boldsymbol\alpha \parallelsum \boldsymbol\beta \parallelsum \boldsymbol\gamma.
\end{align}
This means that all the momentum dependences of the Hamiltonian are factored out as a common factor so that
\begin{align}
\mathcal{H}_{\v k} = (\alpha_x k_x + \alpha_y k_y)(c_0\sigma_0 + \sigma_x + c_y\sigma_y + c_z\sigma_z), \label{eq:non_singular_eff_ham}
\end{align}
where $c_0$, $c_y$, and $c_z$ are the momentum-independent coefficients.
An example for the band structure of this Hamiltonian is shown in Fig.~\ref{fig:eff_bands}(a).
This form of the Hamiltonian always yields the non-singular band touching since its eigenvectors are independent of momentum too.

For the three dimensional case, a general effective Hamiltonian for a linear band touching with a flat band is given by
\begin{align}
\mathcal{H}_{\v k} =& \sum_j \left( b_jk_j\sigma_0 + \alpha_jk_j\sigma_x +  \beta_jk_j\sigma_x + \gamma_jk_j\sigma_x \right),
\end{align}
where $j$ runs from $x$ to $z$.
The flatness condition ($\mathrm{det}\mathcal{H}_{\v k} = 0$) for one of two bands leads to
\begin{align}
b_x^2 =& \alpha_x^2 + \beta_x^2 +\gamma_x^2, \\
b_y^2 =& \alpha_y^2 + \beta_y^2 +\gamma_y^2, \\
b_z^2 =& \alpha_z^2 + \beta_z^2 +\gamma_z^2,
\end{align}
and
\begin{align}
b_xb_y =& \alpha_x\alpha_y + \beta_x\beta_y + \gamma_x\gamma_y, \\
b_yb_z =& \alpha_y\alpha_z + \beta_y\beta_z + \gamma_y\gamma_z, \\
b_xb_z =& \alpha_x\alpha_z + \beta_x\beta_z + \gamma_x\gamma_z.
\end{align}
After the same procedure of the 2D case, we obtain the same condition for these coefficients given by $\boldsymbol\alpha\times\boldsymbol\beta = \boldsymbol\alpha\times\boldsymbol\gamma = \boldsymbol\gamma\times\boldsymbol\beta = 0$
where $\boldsymbol\alpha = (\alpha_x,\alpha_y,\alpha_z)$, $\boldsymbol\beta = (\beta_x,\beta_y,\beta_z)$, and $\boldsymbol\gamma = (\gamma_x,\gamma_y,\gamma_z)$.
This again leads to the same conclusion $\boldsymbol\alpha \parallelsum \boldsymbol\beta \parallelsum \boldsymbol\gamma$ which implies the flat band is non-singular.

\subsection{Quadratic touching}\label{sec:app_quad_touching}
Now, let us consider the quadratic touching of the flat band.
Oshikawa demonstrated that the linear touching between two bands in 3D can be described effectively by
\begin{align}
\tilde{\mathcal{H}}_{\tilde{\mathbf{p}}} = \tilde{p}_z\sigma_z + \tilde{p}_x\sigma_x + \tilde{p}_y\sigma_y + \tilde{\mathbf{p}}\cdot\v b\sigma_0, \label{eq:weyl}
\end{align}
where $\tilde{\mathbf{p}} = \mathcal{T}\mathbf{k}$ with a proper upper triangular matrix $\mathcal{T}$ having positive diagonal elements~\cite{Oshikawa1994}.
In 2D quadratic band touching, the formula (\ref{eq:weyl}) is still useful since we have three independent quadratic terms proportional to $k_x^2$, $k_y^2$, and $k_xk_y$.
By replacing $(k_x,k_y,k_z)$ with $(k_x^2, k_xk_y, k_y^2)$, we obtain a generic form of the quadratic Hamiltonian as follows.
\begin{align}
\tilde{\mathcal{H}}_{\v k} =& (t_1 k_x^2 + t_2 k_x k_y + t_3 k_y^2)\sigma_z + (t_4k_xk_y +t_5k_y^2)\sigma_y \nonumber\\
&+t_6k_y^2\sigma_x + (b_1 k_x^2 + b_2 k_x k_y + b_3 k_y^2)\sigma_0, \label{eq:singular_eff_ham}
\end{align}
where we use the abbreviated notation $t_i$ instead of $\mathcal{T}_{\alpha,\beta}$ for convenience.
Let us denote $h_x(\v k) = t_6k_y^2$, $h_y(\v k) = t_4k_xk_y +t_5k_y^2$, $h_z(\v k) = t_1 k_x^2 + t_2 k_x k_y + t_3 k_y^2$, and $h_0(\v k) = b_1 k_x^2 + b_2 k_x k_y + b_3 k_y^2$.
To have a flat band, following conditions are required
\begin{align}
t_1^2 =& b_1^2, \label{eq:quad_1}\\
t_1t_2 =& b_1 b_2, \label{eq:quad_2}\\
t_3^2 + t_5^2 + t_6^2 =& b_3^2, \label{eq:quad_3}\\
t_2^2 +2 t_1t_3 + t_4^2 =& b_2^2 +2 b_1b_3, \label{eq:quad_4}\\
t_2t_3 + t_4t_5 =& b_2b_3, \label{eq:quad_5}
\end{align}
which are obtained from the condition $\mathrm{det}\tilde{\mathcal{H}}_{\v k}=0$.
For example, the Hamiltonian has a flat band when $t_1 = t_2 = t_3 = t_5 = b_1 = b_2 = 1$, $t_4 = 2$, $t_6 = \sqrt{7}$, and $b_3 = 3$ which is shown in Fig.~\ref{fig:eff_bands}(b).

\subsubsection{$t_1\neq 0$ and $t_4\neq 0$}\label{sec:app_quad_touching_1}
In this case, from (\ref{eq:quad_1}) and (\ref{eq:quad_2}), we obtain $t_2^2 = b_2^2$.
Then, from (\ref{eq:quad_4}) and (\ref{eq:quad_5}), we have
\begin{align}
t_5 =& \frac{t_2}{2t_1}t_4.\label{eq:flatness_cond_s1}
\end{align}
Eliminating $b_3$ in (\ref{eq:quad_3}) by using (\ref{eq:quad_5}), $t_6$ is given by
\begin{align}
t_6^2 = \frac{t_4^2}{4t_1^2} (4t_1t_3 + t_4^2 - t_2^2).\label{eq:flatness_cond_s2}
\end{align}
As a result, we note that independent parameters are just $t_1$, $t_2$, $t_3$, and $t_4$.
One can see that the Hamiltonian has a flat band with the singular touching at $\v k=0$ as is clear from the form of the eigenvector given by
\begin{align}
\v v_{\v k} \propto \bpm -t_6^2k_y^2+i(t_4k_xk_y +t_5k_y^2) \\ h(\v k)+(t_1 k_x^2 + t_2 k_x k_y + t_3 k_y^2) \epm,
\end{align}
which cannot have any common factor as long as $t_1$ and $t_4$ are nonzero.
Here, $h(\v k) = (h_x^2+h_y^2+h_z^2)^{1/2}$.
While this is the conclusion when the flat band is the lower band, we have the same conclusion for the opposite case.

\subsubsection{$t_1 \neq 0$ and $t_4= 0$}\label{app:b}

From $t_2^2 = b_2^2$ and (\ref{eq:quad_5}), we have $t_3^2 = b_2^2$.
Then, from (\ref{eq:quad_3}), we conclude that $t_5=t_6=0$.
Consequently, the eigenvector is of the form
\begin{align}
\v v_{\v k} \propto \bpm 0 \\ h(\v k)+(t_1 k_x^2 + t_2 k_x k_y + t_3 k_y^2) \epm \propto \bpm 0 \\ 1 \epm,
\end{align}
which implies that the quadratic touching of the flat band is non-singular.

\subsubsection{$t_1 = 0$ and $t_4= 0$}\label{app:c}

From $t_1=0$, we have $b_1=0$.
Then, (\ref{eq:quad_4}) reduces to $t_2^2=b_2^2$ because we assume $t_4=0$.
It leads to $t_3^2=b_3^2$ from (\ref{eq:quad_5}).
As a result, from (\ref{eq:quad_3}), we have $t_5=t_6=0$ so that the eigenvector becomes
\begin{align}
\v v_{\v k} \propto \bpm 0 \\ h(\v k)+( t_2 k_x k_y + t_3 k_y^2) \epm \propto \bpm 0 \\ 1 \epm.
\end{align}
This means the quadratic touching of the flat band in this case is also non-singular.

\subsubsection{$t_1 = 0$ and $t_4 \neq 0$}\label{app:d}

In this case, (\ref{eq:quad_4}) becomes $t_2^2+t_4^2=b_2^2$.
Then, from (\ref{eq:quad_3}) and (\ref{eq:quad_5}), removing $b_2$ and $b_3$, we obtain
\begin{align}
0 = (t_2t_5-t_3t_4)^2 + t_6^2(t_2^2+t_4^2), \label{eq:quad_6}
\end{align}
which means $t_2t_5=t_3t_4$ and $t_6=0$ because we assume $t_4\neq 0$.
If $t_2=0$, we have $t_3=0$ due to (\ref{eq:quad_6}) so that the eigenvector becomes
\begin{align}
\v v_{\v k} \propto \bpm i(t_4k_xk_y +t_5k_y^2) \\ h(\v k)\epm \propto \bpm i \\ \pm 1 \epm,
\end{align}
which is non-singular at $\v k=0$.
If $t_2\neq 0$, on the other hand, the eigenvector is of the form
\begin{align}
\v v_{\v k} \propto \bpm iA(t_2k_xk_y +t_3k_y^2) \\ h(\v k)+( t_2 k_x k_y + t_3 k_y^2) \epm,
\end{align}
where the constant $A$ is introduced to reflect the condition $t_2t_5=t_3t_4$.
Then, $h(\v k)=(A^2+1)^{1/2}|t_2 k_x k_y + t_3 k_y^2|$ which reduces the eigenvector into the form
\begin{align}
\v v_{\v k} \propto \bpm iA \\ 1 \pm \sqrt{A^2+1}\epm,
\end{align}
which is non-singular at $\v k=0$.
Note that for all the non-singular touching cases from \textit{b} to \textit{d} above, the Hamiltonian is composed of only a single Pauli matrix.
That is, the general form of the flat band Hamiltonian with the non-singular touching is given by
\begin{align}
\mathcal{H}_{\v k} = (t_1^\prime k_x^2 + t_2^\prime k_x k_y + t_3^\prime k_y^2)(\sigma_z+\sigma_0)
\end{align}
after a proper rotation of the Pauli matrices.

\section{Phase transition properties}\label{app:transition}

Based on the classification of the quadratic touching of a 2D flat band model in the previous section, we investigate the response of a flat band against a generic perturbation of the form $\mathcal{H}_k^\prime(\lambda) = \sum_\alpha f_\alpha(\lambda)\sigma_\alpha $ where $\alpha$ is from $x$ to $z$.
We assume that $f_\alpha(0)=0$, so that we have a flat band at $\lambda =0$.
Since we are interested in the behavior of a flat band in the vicinity of the transition point ($\lambda=0$), we consider $\mathcal{H}_k^\prime(\lambda) \approx \delta\lambda \sum_\alpha \Lambda_\alpha \sigma_\lambda $ where $\Lambda_\alpha = \partial_{\lambda} f_\alpha |_{\lambda = 0}$.
%
%


\subsection{Non-singular touching}

As discussed in the previous section, for a non-singular touching between the flat and quadratic bands, the Hamiltonian including the perturbation is given by
\begin{align}
\mathcal{H}_{\v k} + \mathcal{H}_{\v k}^\prime =& (t_1 k_x^2 + t_2 k_x k_y + t_3 k_y^2)\sigma_z + \delta\lambda\sum_{\alpha=x,y,z}\Lambda_\alpha\sigma_\alpha,
\end{align}
where we omit the term proportional to $\sigma_0$ which is irrelevant in analyzing the phase transition behavior.
First, if $\Lambda_x$ or $\Lambda_y$ is nonzero, there is no band touching except when $\delta\lambda=0$ at which we assume the flat band's quadratic touching.
On the other hand, if $\Lambda_x=\Lambda_y=0$ and $\Lambda_z\neq 0$, the equation $t_1k_x^2 + t_2k_xk_y + t_3k_y^2 +\Lambda_z\delta\lambda = 0$ haa real solutions for $k_x$ and $k_y$.
For $k_y$, we have
\begin{align}
k_y = \frac{-t_2k_x \pm \sqrt{(t_2^2-4t_1t_3)k_x^2 - 4\Lambda_zt_3\delta\lambda}}{2t_3},
\end{align}
which becomes real-valued when $(t_2^2-4t_1t_3)k_x^2 > 4\Lambda_zt_3\delta\lambda$.
If $t_2^2-4t_1t_3 > 0$, the inequality always holds for the sufficiently large $k_x$ or small $\delta\lambda$.
Since there is only one constraint for $k_x$ and $k_y$, we have a transition between two line touching semimetals through the intermediate quadratic point touching of the flat band.
However, if $t_2^2-4t_1t_3 < 0$, we can have an insulating phase for $\delta\lambda <0$ by assuming $\Lambda_1t_3 < 0$.
After the quadratic touching at $\delta\lambda=0$, we have a line touching semimetal for $\delta\lambda >0$.
This is the insulator-to-metal transition.

\subsection{singular touching}

First, let us consider the case where $t_6 = 0$ in (\ref{eq:singular_eff_ham}).
If $\Lambda_x\neq 0$, we always have an insulating phase for $\delta\lambda \neq 0$ which means the insulator-to-insulator transition.
On the other hand, if $\Lambda_x = 0$ but $\Lambda_y$ or $\Lambda_z$ are nonzero, the band touching requires real solutions of following two equations.
\begin{align}
&t_1k_x^2 + t_2k_xk_y + t_3k_y^2 +\Lambda_z\delta\lambda = 0,\label{eq:singular_eq_5} \\
&t_4k_xk_y + t_5k_y^2 + \Lambda_y\delta\lambda = 0.\label{eq:singular_eq_6}
\end{align}
Without loss of the generality, we set $t_1=1$.
Combining two equations in the above and applying the identity $t_5=t_2t_4/2t_1 = t_2t_4/2$ from the flatness condition (\ref{eq:flatness_cond_s1}), we obtain
\begin{align}
&t_4\left( k_x^2 + \left( t_3- \frac{t_2^2}{2}\right)k_y^2\right) +\delta\lambda\left( t_4\Lambda_z - t_2\Lambda_y \right) = 0, \label{eq:singular_eq_1}\\
&t_4\left( k_xk_y + \frac{t_2}{2}k_y^2\right) + \delta\lambda\Lambda_y = 0.\label{eq:singular_eq_2}
\end{align}
From (\ref{eq:singular_eq_1}), we have $k_x = \pm\sqrt{g_1 - g_2k_y^2}$ where $g_1 = -\delta\lambda (\Lambda_z - t_2\Lambda_y/t_4)$ and $g_2 =t_3-t_2^2/2$.
Plugging this into (\ref{eq:singular_eq_2}), we obtain a real solution of $k_y$ from
\begin{align}
k_y^2 = \frac{2}{t_4^2}\left( \Lambda_z \pm \sqrt{\Lambda_z^2 + \Lambda_y^2}\right)\delta \lambda, \label{eq:singular_eq_3}
\end{align}
where we take the plus (minus) sign for the positive (negative) $\delta\lambda$. 
For $k_x$ to be also real-valued, $g_1-g_2k_y^2$ should be positive.
We check the inequality $g_1-g_2k_y^2 > 0$ by replacing $k_y^2$ in $g_1-g_2k_y^2$ with (\ref{eq:singular_eq_3}) which leads to
\begin{align}
g_1-g_2k_y^2 = \frac{\delta\lambda}{t_4^2}\left( 2t_3\Lambda_z +t_2t_4\Lambda_y \pm (t_2^2-2t_3)\sqrt{\Lambda_y^2+\Lambda_z^2} \right), \label{eq:singular_eq_4}
\end{align}
where we have used the relation $t_4^2 = t_2^2-4t_3$ obtained
 from (\ref{eq:flatness_cond_s2}) with $t_6=0$.
Note that $t_2^2-2t_3$ is always positive.
Let us denote $C_1=(2t_3\Lambda_z +t_2t_4\Lambda_y)^2$ and $C_2 = (t_2^2-2t_3)^2(\Lambda_y^2+\Lambda_z^2)$.
Then, one can show (\ref{eq:singular_eq_4}) is positive because
\begin{align}
C_2 -C_1 = (t_2t_4\Lambda_z - 2t_3\Lambda_y)^2.
\end{align}
This result implies that we always have a band touching regardless of the sign of $\delta\lambda$ when $t_6=0$ and $\Lambda_x=0$.
Furthermore it is point touching unlike the non-singular case because it is obtained from the two independent constraints (\ref{eq:singular_eq_5}) and (\ref{eq:singular_eq_6}) for two variables $k_x$ and $k_y$.
That is, we only have the phase transition from a point-node semimetal to another point-node semimetal through the quadratic flat band touching.
However, we cannot obtain the insulator-to-metal transition in this case.

Second, if $t_6\neq 0$, we have one more constraint
\begin{align}
t_6k_y^2 = -\delta\lambda\Lambda_x,
\end{align}
in addition to (\ref{eq:singular_eq_5}) and (\ref{eq:singular_eq_6}).
As a result, for the nonzero $\Lambda_x$, we can have the insulating phase when $\delta\lambda <0 $ if $\Lambda_x/t_6 < 0$.
In this case we have a point-node semimetal for $\delta\lambda >0$ if
\begin{align}
&t_6^2(\Lambda_x^2+\Lambda_y^2+\Lambda_z^2) = \left( t_6\Lambda_z + \frac{t_4\Lambda_x}{2}\right)^2,
\end{align}
and an insulating phase otherwise.
That is, in $t_6\neq 0$ case, the insulator-to-metal phase transition is allowed.

%
%
%
%
%

\section{Chern number of nearly flat bands}\label{sec:chern}

In general, a gapped 2 by 2 Hamiltonian is written in the form $\mathcal{H} = \sum_{\alpha=x,y,z} d_\alpha(k_x,k_y)\sigma_\alpha$, where $d_\alpha(k_x,k_y)$ is a real-valued function and $\sigma_\alpha$ is the Pauli matrix.
Then, the corresponding Chern number of the occupied band is given by
\begin{align}
\nu = \frac{1}{2\pi}\int d^2\v k \mathcal{F}_{xy},
\end{align}
where the Berry curvature $\mathcal{F}_{xy}$ is defined as
\begin{align}
\mathcal{F}_{xy} = \frac{1}{2}\varepsilon_{\alpha\beta\gamma} \hat{d}_\alpha \partial_x\hat{d}_\beta \partial_y\hat{d}_\gamma.
\end{align}
Here, $\hat{d}_\alpha = d_\alpha/( d_x^2 +d_y^2 +d_z^2 )^{1/2}$, and $\varepsilon_{\alpha\beta\gamma}$ is the Levi-Civita tensor.

We calculate the Chern number of the nearly flat band obtained by the gap opening process of the singular flat band model given by
\begin{align}
\mathcal{H}_{\v k} = \frac{k_x^2 - k_y^2}{2}\sigma_z + k_xk_y\sigma_y + \frac{k_x^2 + k_y^2}{2}\sigma_0,
\end{align} 
where $\sigma_0$ is the identity matrix.
We examine three basic perturbations $\mathcal{H}^{(x)}= m\sigma_x$, $\mathcal{H}^{(y)}= m\sigma_y$ and $\mathcal{H}^{(z)}= m\sigma_z$.
First, $\mathcal{H}^{(x)}$ can gap out the singular touching, and the Berry curvature of the separate nearly flat band becomes
\begin{align}
\mathcal{F}_{xy}^{(x)} = -\frac{4mk^2}{(4m^2 + k^4)^{\frac{3}{2}}},
\end{align}
where $k^2 = k_x^2 + k_y^2$.
Then, the Chern number is obtained as
\begin{align}
\nu_m = -\mathrm{sgn}(m),
\end{align}
which leads to $\Delta\nu = \nu_+ - \nu_- = 2$.
While the Chern number of a continuum model depends on the regularization scheme, the finite Chern number difference means that we have a nonzero Chern number at least when $m<0$ or $m>0$.
On the other hand, the other perturbations $\mathcal{H}^{(y)}$ and $\mathcal{H}^{(z)}$ cannot open a gap.
Instead, the quadratic touching at $\v k=0$ is split into two linear crossings.
The crossing points are located at $\v k = \pm(\sqrt{|m|},-\mathrm{sgm}(m)\sqrt{|m|})$ for $\mathcal{H}^{(y)}$, $\v k = \pm(0,\sqrt{2m})$ for $\mathcal{H}^{(z)}$ with positive $m$, and $\v k = \pm(\sqrt{-2m},0)$ for $\mathcal{H}^{(z)}$ with negative $m$.
%

%

\section{Lieb lattice}\label{sec:lieb}

While the doubly degenerate band touching is quite generic among the flat band models, we sometimes encounter with the higher degeneracy as in the Lieb lattice (Fig.~\ref{fig:lieb}(a)) discussed in this section.
The Hamiltonian of the Lieb lattice is given by
\begin{align}
\mathcal{H}_{\v k} = \bpm 0 & 1+e^{ik_x} & 0 \\ 1+e^{-ik_x} & 0 & 1+e^{-ik_y} \\ 0 & 1+e^{ik_y}& 0 \epm,
\end{align}
which has a flat band at the zero energy, and the upper and lower bands, described by $E_\pm(\v k) = \pm (4+2\cos k_x +2\cos k_y)^{1/2}$, touch linearly with each other at the zero energy at $\v k=(\pi,\pi)$ as shown in Fig.~\ref{fig:lieb}(e).
The eigenvector of the flat band is given by
\begin{align}
\v v_{\v k} = \frac{1}{\sqrt{E_+(\v k)}}\bpm 1+e^{-ik_y} \\ 0 \\ -1-e^{-ik_x} \epm,
\end{align}
which has the immovable discontinuity at $\v k=(\pi,\pi)$.
The relevant CLS is described by
\begin{align}
\mathbf{A}_{0,R} = \frac{1}{2} \bpm \delta^{\v R}_{(0,0)}+\delta^{\v R}_{(0,1)} \\ 0 \\ -\delta^{\v R}_{(0,0)}-\delta^{\v R}_{(1,0)} \epm, \label{eq:lieb_cls_0}
\end{align}
which is obtained by choosing $\alpha_{\v k} = E_+(\v k)^{1/2}$.
This is shown in Fig.~\ref{fig:lieb}(a).

\section{Modified Lieb lattice}\label{sec:modified_lieb}

%
We can move the Dirac point to $\v k = (0,0)$ by changing the signs of the inter-unit cell hopping processes of the Lieb lattice model as plotted in Fig.~\ref{fig:lieb}(b).
We call it the modified Lieb lattice model.
The Hamiltonian is given by
\begin{align}
\mathcal{H}_{\v k} = \bpm 0 & 1-e^{ik_x} & 0 \\ 1-e^{-ik_x} & 0 & 1-e^{-ik_y} \\ 0 & 1-e^{ik_y}& 0 \epm,
\end{align}
which has a flat band at the zero energy, and the upper and lower bands, $E_\pm(\v k) = \pm (4-2\cos k_x -2\cos k_y)^{1/2}$, show a linear crossing at the zero energy at $\v k=(0,0)$ as plotted in Fig.~\ref{fig:lieb}(e).
The eigenvector of the flat band is obtained as
\begin{align}
\v v_{\v k} = \frac{1}{\sqrt{E_+(\v k)}}\bpm -1+e^{-ik_y} \\ 0 \\ 1-e^{-ik_x} \epm,
\end{align}
which is discontinuous at $\v k=(0,0)$.
From this, we obtain the corresponding CLS of the form
\begin{align}
\mathbf{A}_{0,R} = \frac{1}{2} \bpm -\delta^{\v R}_{(0,0)}+\delta^{\v R}_{(0,1)} \\ 0 \\ \delta^{\v R}_{(0,0)}-\delta^{\v R}_{(1,0)} \epm, \label{eq:lieb_cls_1}
\end{align}
from $\alpha_{\v k} = E_+(\v k)^{1/2}$.
This is described in Fig.~\ref{fig:lieb}(b).
The discontinuity of the eigenvector at $\v k =(0,0)$ implies $N$ translated copies of the CLS in the above are incomplete.

\begin{figure}
	\begin{center}
		\includegraphics[width=1\columnwidth]{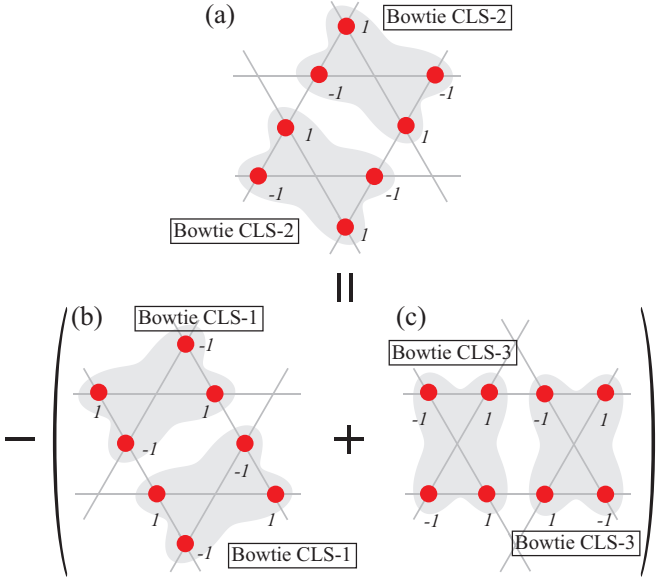}
	\end{center}
	\caption{ The sum of the six bowtie CLSs around a hexagon vanishes. In other words, the combination of two bowtie CLSs in (a) is a linear combination of bowtie CLS-1s and bowtie CLS-3s in (b) and (c) respectively.
	}
	\label{fig:cls_2}
\end{figure}

\begin{figure*}
	\begin{center}
		\includegraphics[width=2\columnwidth]{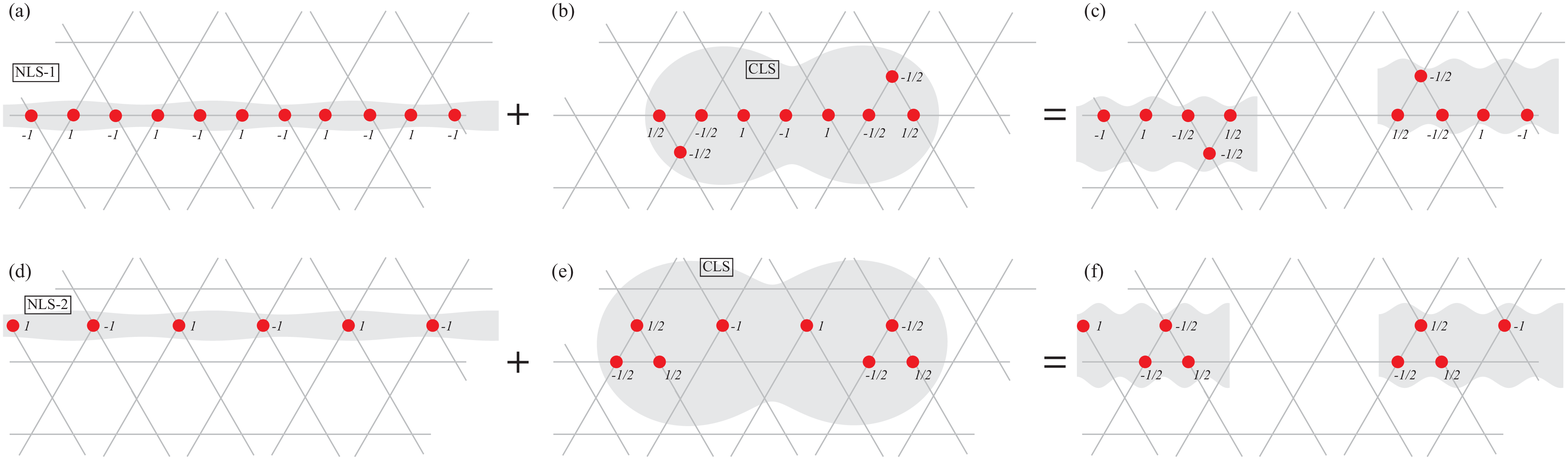}
	\end{center}
	\caption{ The NLS-1 and NLS-2 of the kagome-3 model in (a) and (d) are disconnected into two pieces as shown in (c) and (f) by adding CLSs in (b) and (e) to them respectively.
	}
	\label{fig:kagome_3_nls}
\end{figure*}

\section{Kagome-3 model}\label{app:kagome_3}

In this sections, we consider the kagome-3 model studied by D. L. Bergman \textit{et al}~\cite{Balents2008}.
While they claimed that there are four NLSs, we show that their NLSs are actually contractible.
According to our theory, the flat bands of the kagome-3 model are actually non-singular and one can find the proper CLSs that form a complete set spanning the flat bands.

The elements of the Hamiltonian matrix of the kagome-3 model is given by
\begin{align}
\mathcal{H}_{jj} &= e^{i\mathbf{k}\cdot\mathbf{a}_j} + e^{-i\mathbf{k}\cdot\mathbf{a}_j}, \\
\mathcal{H}_{21} &= 1 + e^{-i\mathbf{k}\cdot\mathbf{a}_1} + e^{-i\mathbf{k}\cdot\mathbf{a}_2} + e^{i\mathbf{k}\cdot\mathbf{a}_3}, \\
\mathcal{H}_{31} &= 1 + e^{i\mathbf{k}\cdot\mathbf{a}_1} + e^{-i\mathbf{k}\cdot\mathbf{a}_2} + e^{i\mathbf{k}\cdot\mathbf{a}_3}, \\
\mathcal{H}_{32} &= 1 + e^{i\mathbf{k}\cdot\mathbf{a}_1} + e^{-i\mathbf{k}\cdot\mathbf{a}_2} + e^{-i\mathbf{k}\cdot\mathbf{a}_3},
\end{align}
where $\mathbf{a}_1 = a\hat{x}$, $\mathbf{a}_2 = -1/2\hat{x} +\sqrt{3}/2\hat{y}$, and $\mathbf{a}_3 = -1/2\hat{x} -\sqrt{3}/2\hat{y}$ illustrated in Fig.~\ref{fig:kagome_3}(b).
We assume $a=1$ for simplicity.
The eigenvalues and eigenvectors are evaluated as
\begin{align}
E^{(1)}_\mathbf{k} &= E^{(2)}_\mathbf{k} = -2, \\
E^{(3)}_\mathbf{k} &= 4 + 2\cos k_x + 4\cos\frac{k_x}{2}\cos\frac{\sqrt{3}k_y}{2},
\end{align}
and
\begin{align}
\mathbf{v}^{(1)}_\mathbf{k} &= c_1 \bpm -1 - e^{-i\mathbf{k}\cdot\mathbf{a}_3} \\ 0  \\ 1 + e^{i\mathbf{k}\cdot\mathbf{a}_1} \epm, \\
\mathbf{v}^{(2)}_\mathbf{k} &= c_2 \bpm -e^{i\mathbf{k}\cdot\mathbf{a}_1} - e^{-i\mathbf{k}\cdot\mathbf{a}_3} \\ 1 + e^{i\mathbf{k}\cdot\mathbf{a}_1} \\ 0 \epm,
\end{align}
for two degenerate flat bands where $c_1$ and $c_2$ are normalization coefficients.
One can quickly check that $\mathbf{v}^{(1)}_\mathbf{k}/c_1$ and $\mathbf{v}^{(2)}_\mathbf{k}/c_2$ correspond to the CLSs called the bowtie CLS-1 and -2 respectively as illustrated in Fig.~\ref{fig:kagome_3}(a).
Combining $\mathbf{v}^{(1)}_\mathbf{k}$ and $\mathbf{v}^{(2)}_\mathbf{k}$, we can find another form of the eigenvector
\begin{align}
\mathbf{v}^{(3)}_\mathbf{k} &= c_3 \bpm 0 \\  1 + e^{-i\mathbf{k}\cdot\mathbf{a}_3}  \\ -1 - e^{-i\mathbf{k}\cdot\mathbf{a}_2} \epm.
\end{align}
The relevant CLS is denoted by the bowtie CLS-3 and depicted in Fig.~\ref{fig:kagome_3}(a).
Note that although $\mathbf{v}^{(3)}_\mathbf{k}$ is constructed from $\mathbf{v}^{(1)}_\mathbf{k}$ and $\mathbf{v}^{(2)}_\mathbf{k}$, it does not mean that one can obtain a single bowtie CLS-3 from the bowtie CLS-1 and -2.
As described in Fig.~\ref{fig:cls_2}, one can show that a couple of the bowtie CLSs of one kind can be constructed from other kinds of bowtie CLSs.
However, it is impossible to represent a single bowtie CLS by the linear combination of other kinds of bowtie CLSs.

Bergman \textit{et al} showed that two sets of $N$ translated copies of the bowtie CLS-1 and -2 are incomplete and suggested that there must be four NLSs plotted in Fig.~ \ref{fig:kagome_3}(b).
It is correct that their bowtie CLSs form incomplete sets as manifested by the singularities in $\v v^{(1)}_{\v k}$ and $\v v^{(2)}_{\v k}$ at $\v k = (\pi,\pi/\sqrt{3})$ and $\v k = (\pi,-\pi/\sqrt{3})$ respectively.
However, their NLSs are not independent of the bowtie CLSs as shown follows.
First, we show that the NLS-1 can be constructed by the sum of the translated copies of bowtie CLS-1 and -2.
The CLSs corresponding to $\v v^{(1)}_{\v k}$ and $\v v^{(2)}_{\v k}$ are given by
\begin{align}
\v A^{(1)}_{\v R^\prime,\v R} = \frac{1}{2} \bpm -\delta_{\v R - \v R^\prime} - \delta_{\v R - \v R^\prime - \v a_3} \\ 0 \\ \delta_{\v R - \v R^\prime} + \delta_{\v R - \v R^\prime + \v a_1} \epm, \\
\v A^{(2)}_{\v R^\prime,\v R} = \frac{1}{2} \bpm -\delta_{\v R - \v R^\prime} - \delta_{\v R - \v R^\prime + \v a_2} \\  \delta_{\v R - \v R^\prime} + \delta_{\v R - \v R^\prime - \v a_1} \\ 0 \epm.
\end{align}
Then the NLS-1 and -2, denoted by $\v B_{\v R}^{(1)}$ and $\v B_{\v R}^{(2)}$, are represented as
\begin{align}
\v B_{\v R}^{(1)} &= \sum_{n=1}^{N_x} \left( A^{(1)}_{\v R^\prime + n\v a_1,\v R} - A^{(2)}_{\v R^\prime+ n\v a_1,\v R} \right), \\
\v B_{\v R}^{(2)} &= 2\sum_{n=1}^{N_x/2} \left( A^{(1)}_{\v R^\prime + 2n\v a_1,\v R} - A^{(2)}_{\v R^\prime+ (2n-1)\v a_1,\v R} \right) -\v B_{\v R}^{(1)}
\end{align}
where $N_x$ is the system size along $x$ direction.
This shows that two NLSs suggested by Bergman \textit{et al} are not non-contractible, and can be disconnected by adding finite number of CLSs as described in Fig.~\ref{fig:kagome_3_nls}.
In Fig.~\ref{fig:kagome_3_nls}, the CLSs in (b) and (e) are obtained from the combinations of the bowtie CLS-1s and -2s at different positions.

Then, are there any genuine NLSs?
Our answer is that the flat bands of the kagome-3 model are actually non-singular and we do not need any NLSs.
The crucial point is that although there are singular momenta in $\v v^{(1)}_{\v k}$ and $\v v^{(2)}_{\v k}$, we can recombine these two eigenvectors to obtain non-singular set of eigenvectors because two flat bands are completely degenerate.
These non-singular eigenvectors are obtained as
\begin{align}
\v w^{(1)}_{\v k} &\propto \frac{\v v^{(1)}_{\v k}}{c_1} - \frac{\v v^{(2)}_{\v k}}{c_2} = \bpm -1+e^{i\v k\cdot\v a_1} \\ -1-e^{i\v k\cdot\v a_1} \\ 1+e^{i\v k\cdot\v a_1} \epm , \\
\v w^{(2)}_{\v k} &\propto \frac{\v v^{(1)}_{\v k}}{c_1} + \frac{\v v^{(2)}_{\v k}}{c_2} = \bpm -1-e^{i\v k\cdot\v a_1}-2e^{-i\v k\cdot\v a_3} \\ 1+e^{i\v k\cdot\v a_1} \\ 1+e^{i\v k\cdot\v a_1} \epm.
\end{align}
The CLSs corresponding to these are plotted in Fig.~\ref{fig:kagome_3}(c).
%


\begin{thebibliography}{72}
\bibitem{note1} The topology of the NLS of the flat band in real space is completely different from the conventional topology of the usual topological phases defined in momentum space.
\end{thebibliography}





\end{document}